\documentclass[12pt,preprint]{aastex}

\newcommand{\etal  }{{et al.} }
\newcommand{\msun}{\thinspace M_\odot}  
\newcommand{\rsun}{\thinspace R_\odot}  
\newcommand{\vect}[1]{\mbox{\boldmath$#1$}}

\newcommand{\rhoc}{\rho_{\rm c}}
\newcommand{\nc}{n_{\rm c}}

\newcommand{\ceta}{c_{\eta}}

\newcommand{\cm  }{\,{\rm cm}^{-3} } 
\newcommand{\jcm}{{\rm cm^2\,s^{-1}}}
\newcommand{\dfrac}[2]{{\displaystyle \frac{#1}{#2}}  }
\newcommand{\km  }{\,{\rm km\, s^{-1}} } 
\newcommand{\kg  }{\,{\rm kG} } 
\newcommand{\betap }{\beta_{\rm p}}

\shorttitle{Magnetic Fields and Rotations of Protostars}
\shortauthors{Machida  \etal 2006}
\begin{document}
\title{Magnetic Fields and Rotations of Protostars}

\author{Masahiro N. Machida\altaffilmark{1} and Shu-ichiro Inutsuka\altaffilmark{1}, and Tomoaki Matsumoto\altaffilmark{2}} 
\altaffiltext{1}{Department of Physics, Graduate School of Science, Kyoto University, Sakyo-ku, Kyoto 606-8502, Japan; machidam@scphys.kyoto-u.ac.jp, inutsuka@tap.scphys.kyoto-u.ac.jp}
\altaffiltext{2}{Faculty of Humanity and Environment, Hosei University, Fujimi, Chiyoda-ku, Tokyo 102-8160, Japan; matsu@i.hosei.ac.jp}

\begin{abstract}
The early evolution of the magnetic field and angular momentum of 
 newly formed protostars are studied using three-dimensional 
 resistive MHD nested grid simulations. 
Starting with a Bonnor--Ebert isothermal cloud rotating in a uniform magnetic field, we calculate the cloud evolution from the molecular cloud core ($\nc\simeq10^4 \cm$, $r=4.6\times 10^5$\,AU, where $\nc$ and $r$ are central density and radius, respectively) to the stellar core ($\nc\simeq10^{22} \cm$, $r \sim 1 R_\odot$). 
The magnetic field strengths at the center of clouds with the same initial angular momentum but different magnetic field strengths converge to a certain value as the clouds collapse for $\nc \lesssim 10^{12}\cm$.
For $10^{12}\lesssim \nc \lesssim 10^{16}\cm$, Ohmic dissipation largely removes the magnetic field from a collapsing cloud core, and the magnetic field lines, which are strongly twisted for $\nc \lesssim 10^{12}\cm$, are de-collimated.
The magnetic field lines are twisted and amplified again for $\nc\gtrsim 10^{16}\cm$, because the magnetic field is recoupled with warm gas. 
Finally,  protostars at their formation epoch ($\nc \simeq 10^{21}\cm$) have magnetic fields of $\sim$0.1--1\,kG, which is comparable to observations.
The magnetic field strength of a protostar depends slightly on the angular momentum of the host cloud.
A protostar formed from a slowly rotating cloud core has a stronger magnetic field.
The evolution of the angular momentum is closely related to the evolution of the magnetic field.
The angular momentum in a collapsing cloud is removed by magnetic effects such as magnetic braking, outflow and jets.   
The formed protostars have rotation periods of 0.1--2\,days at their formation epoch, which is slightly shorter than observations.
This indicates that a further removal mechanism for the angular momentum, such as interactions between the protostar and the disk, wind, or jets, is important in the further evolution of protostars. 
\end{abstract}

\keywords{ISM: clouds: ISM: magnetic fields---MHD---stars: formation---stars: rotation}

\section{Introduction}
The Lorentz and centrifugal forces play important roles in the star formation process.
While gravity and thermal pressure are isotropic forces, the Lorentz and centrifugal forces are anisotropic forces and are closely related to disk formation, outflow, and jets in collapsing clouds.
Molecular clouds have $\sim 2\%$ rotational energy against gravitational energy \citep{goodman93, caselli02},
while the magnetic energy is comparable to the gravitational energy \citep{crutcher99}.
By considering the conservation of magnetic flux and angular velocity, it is seen that the rotation and magnetic field in a cloud gradually increase as the cloud collapses. 
However,  the magnetic field strength and angular velocity of observed protostars indicate that neither the magnetic flux nor angular momentum is conserved in collapsing clouds.
In general, these anomalies are called the `magnetic flux problem' and `angular momentum problem.' 
The former problem refers to the fact that the magnetic flux of a molecular cloud is much larger than that of a protostar with equivalent mass.
The latter problem is that the specific angular momentum of a molecular cloud is much larger than that of a protostar.
These problems imply that there must be mechanisms removing magnetic flux and angular momentum from a cloud core.
In a collapsing cloud, these two problems are mutually related.
Namely, the angular momentum is removed by magnetic effects (i.e., magnetic braking, outflow, and jets), while the magnetic field is amplified by the shearing motion caused by cloud rotation.
Hence, the magnetic field and rotation cannot be treated independently in considering the magnetic flux and angular momentum problems.
In addition, it is difficult to treat the evolution of the magnetic field and rotation analytically since the density ($n \sim 10^4 \cm$) and scale ($R \sim 10^4$\,AU) of molecular clouds are very different from those of protostars ($n \sim 10^{24}\cm$, $R\sim 1\rsun$). 
Therefore, numerical simulation is needed to study the magnetic field and rotation of protostars formed from molecular clouds.

Angular momentum is removed from a collapsing cloud by magnetic braking and outflow, as shown by \citet{basu94}, \citet{tomisaka02}, and \citet{machida05a}.
\citet{machida05a} found that about 70\% of the total angular momentum is removed from the collapsing cloud core in the isothermal phase ($n \lesssim 10^{11}\cm$).
\citet{tomisaka02} shows that $\sim$99\% of the total angular momentum is transferred by outflow in the adiabatic phase ($10^{11}\cm\lesssim n \lesssim 10^{16}\cm$).
Thus, previous studies show that angular momentum is effectively transferred by magnetic effects.
In contrast, there are relatively few studies of the evolution and removal process of magnetic flux in collapsing cloud cores, especially for high density ($n \gtrsim 10^{12}\cm$).
\citet{nakano02} analytically investigated the dissipation process of the magnetic field in a collapsing cloud core and showed that it is expected that the magnetic flux is removed from the cloud core largely for $10^{12} \cm \lesssim n \lesssim 10^{16}\cm$.

To investigate the evolution of the magnetic field and rotation (or the magnetic flux and angular momentum problems), it is necessary to calculate the cloud evolution of a cloud from the molecular cloud core ($n\simeq 10^4\cm$) to protostar formation ($n\sim 10^{24}\cm$) while accounting for the magnetic field and rotation.
The evolution of a magnetized cloud up to a central density of $n\simeq 10^{15}\cm$ has been investigated by \citet{hosking04}, \citet{ziegler05}, \citet{matsu04}, and \citet{machida04,machida05a,machida05b}.
\citet{tomisaka02} and \citet{banerjee06} calculated cloud evolution up till the formation of a protostar ($n \sim 10^{21}\cm$), and showed that the magnetic field plays important roles in the star-formation process.
 However, they adopted an ideal MHD approximation, which is valid in low-density gas regions ($n \lesssim 10^{12}\cm$), but is not valid in high-density gas regions ($n \gtrsim 10^{12}\cm$).
A significant magnetic flux loss occurs for $10^{12} \cm \lesssim n \lesssim 10^{15}\cm$ by Ohmic dissipation \citep{nakano02} and hence these studies overestimate the magnetic flux in a collapsing cloud, especially in high-density gas regions.

Following \citet{machida06b}, we report detailed results of three-dimensional non-ideal MHD nested grid simulations.
 In this paper, we calculate the cloud evolution from the cloud core ($n_c \simeq 10^4\cm$, $r_c = 4.6 \times 10^5$\,AU) to protostar formation ($n_c \simeq 10^{22} \cm$, $r_c \simeq 1 \rsun$) and discuss the magnetic field and angular momentum of the formed protostar.
 The structure of the paper is as follows. 
 The frameworks of our models are given in \S 2 and the numerical method of our computations is shown in \S3.
 The numerical results are presented in \S 4.  
 We discuss the magnetic field and angular momentum of a protostar and compare our results with observations in \S 5.

\section{Model}
\subsection{Basic Equations}
To study the cloud evolution, we use the three-dimensional resistive MHD nested grid code. 
We solve the resistive MHD equations, including self-gravity:  
\begin{eqnarray} 
& \dfrac{\partial \rho}{\partial t}  + \nabla \cdot (\rho \vect{v}) = 0, & \\
& \rho \dfrac{\partial \vect{v}}{\partial t} 
    + \rho(\vect{v} \cdot \nabla)\vect{v} =
    - \nabla P - \dfrac{1}{4 \pi} \vect{B} \times (\nabla \times \vect{B})
    - \rho \nabla \phi, & \\ 
& \dfrac{\partial \vect{B}}{\partial t} = 
   \nabla \times (\vect{v} \times \vect{B}) + \eta \nabla^2 \vect{B}, & 
\label{eq:reg}\\
& \nabla^2 \phi = 4 \pi G \rho, &
\end{eqnarray}
 where $\rho$, $\vect{v}$, $P$, $\vect{B} $, $\eta$, and $\phi$ denote the density, 
velocity, pressure, magnetic flux density, resistivity, and gravitational potential, respectively. 
The last term in equation (3) denotes the Ohmic dissipation.
Although the dissipation term is expressed exactly by $- \nabla \times (\eta \nabla \times \mbox{\boldmath$B$})$, we simplify it as shown in the equation (3).  This simplification allows us to reduce computational costs considerably, and we can follow the collapse of the cloud up to the stages of the stellar core formation.  
The deviation due to this simplification is examined by test calculations of typical models; this simplification does not affect evolution of the cloud considerably,  and the magnetic field strength deviates by a factor of two at most at the central region.

 To mimic the temperature evolution calculated by \citet{masunaga00}, we adopt the piece-wise polytropic equation of state: 
\begin{equation} 
P = \left\{
\begin{array}{ll}
 c_{s,0}^2 \rho & \rho < \rho_c, \\
 c_{s,0}^2 \rho_c \left( \frac{\rho}{\rho_c}\right)^{7/5} &\rho_c < \rho < \rho_d, \\
 c_{s,0}^2 \rho_c \left( \frac{\rho_d}{\rho_c}\right)^{7/5} \left( \frac{\rho}{\rho_d} \right)^{1.1}
 & \rho_d < \rho < \rho_e, \\
 c_{s,0}^2 \rho_c \left( \frac{\rho_d}{\rho_c}\right)^{7/5} \left( \frac{\rho_e}{\rho_d} \right)^{1.1}
 \left( \frac{\rho}{\rho_e}   \right)^{5/3}
 & \rho > \rho_e, 
\label{eq:eos}
\end{array}
\right.  
\end{equation}
 where $c_{s,0} = 190$\,m\,s$^{-1}$, 
$ \rho_c = 3.84 \times 10^{-13} \, \rm{g} \, \cm$ ($n_c = 10^{11} \cm$), 
$ \rho_d = 3.84 \times 10^{-8} \, \rm{g} \, \cm$  ($n_d =  10^{16} \cm$), and
$ \rho_e = 3.84 \times 10^{-3} \, \rm{g} \, \cm$  ($n_e = 10^{21} \cm$).
For convenience, we define `the protostar formation epoch' as that at which the central density reaches  $\nc = 10^{21} \cm$.
We also call the period for which $n_c < 10^{11}\cm$ `the isothermal phase',  the period for which $10^{11}\cm < \nc < 10^{16}\cm$ `the adiabatic phase', the period for which $10^{16}\cm < \nc < 10^{21}\cm$ `the second collapse phase', and the period for which $\nc > 10^{21} \cm$ `the protostellar phase.'
We, therefore, ignored the effect from heating by Ohmic resistivity, because we adopted a simple equation of state.
Plasma beta in the collapsing clouds maintains $\betap>1$ (see \S4).
Thus, even if all the magnetic field energy is converted into the thermal energy, the gas temperature increases by a factor of two.
However, since plasma beta is  $\betap \gtrsim 10$ in any model when Ohmic dissipation becomes effective (see \S4.1), we can safely neglect the heating effect by Ohmic resistivity.

 In this paper, we adopt a spherical cloud with a critical Bonnor--Ebert \citep{ebert55, bonnor56} density profile $\rho_{\rm BE}$  as the initial condition.
 Initially, the cloud rotates rigidly with angular velocity $\Omega_0$ around the $z$-axis and has a uniform magnetic field $B_0$ parallel to the $z$-axis (or rotation axis).
 To promote contraction, we increase the density by a factor $f$ (density enhancement factor), as 
\begin{eqnarray}
\rho(r) = \left\{
\begin{array}{ll}
\rho_{\rm BE}(r) \, f & \mbox{for} \; \; r < R_{c}, \\
\rho_{\rm BE}(R_c)\, f & \mbox{for}\; \;  r \ge R_{c}, \\
\end{array}
\right. 
\end{eqnarray}
 where $r$ and $R_c$ denote the radius and the critical radius for a Bonnor--Ebert sphere, respectively.
We adopt density enhancement factors of $f=1.2$ and 1.4 (see Table~\ref{table:init})
\footnote{
The density enhancement factor is related to the stability of the initial cloud. 
The cloud is more unstable with larger $f$.
Comparatively stable clouds are considered in this paper.
However, cloud evolution hardly depends on $f$ (for details see \citealt{machida06a}).
}.
  We assume $\rho_{\rm BE}(0) = f\times 3.841 \times 10^{-20}\, {\rm g} \,\cm$, which corresponds to a central number density of $n_{c,0} = f\times10^4\cm$.
Thus, the critical radius of a Bonnor--Ebert sphere $R_c = 6.45\, c_{s,0} [4\pi G \rho_{BE}(0)]^{-1/2}$ corresponds to $ R_c = 4.58 \times 10^4$\,AU for our settings. 
 The initial model is characterized by three nondimensional parameters: $\alpha$, $\omega$, and $\ceta$.
The magnetic field strength and rotation rate are scaled using a central density $\rho_0 = \rho_{\rm BE}(0) f$ as
\begin{equation}
\alpha =  B_0^2 / (4\pi \, \rho_0 \, c_{s,0}^2),
\label{eq:alpha}
\end{equation}
\begin{equation}
\omega = \Omega_0/(4 \pi\,  G \, \rho_0  )^{1/2}.
\label{eq:omega}
\end{equation}
The parameter $\ceta$ represents the degree of resistivity (for details, see the next section).
We calculated 33 models by combining these parameters, which are listed in Table~\ref{table:init}.
The model parameters $\alpha$, $\omega$, and $\ceta$; density enhancement factor $f$; magnetic field $B_0$;  angular velocity $\Omega_0$; total mass $M$ inside the critical radius ($r<R_{\rm c}$); and the ratio of the thermal,  rotational,  and magnetic energies to the gravitational energy $\alpha_0$, $\beta_0$, and $\gamma_0$
are summarized in this table
\footnote{
Denoting the thermal, rotational, magnetic, and gravitational energies as $U$, $K$, $M$, and $W$, the relative factors against the gravitational energy are defined as $\alpha_0 = U/|W|$, $\beta_0 = K/|W|$, and $\gamma_0 = M/|W|$.
}.

\subsection{Resistivity and Magnetic Reynolds Number}
\begin{figure}
\begin{center}
\includegraphics[width=150mm]{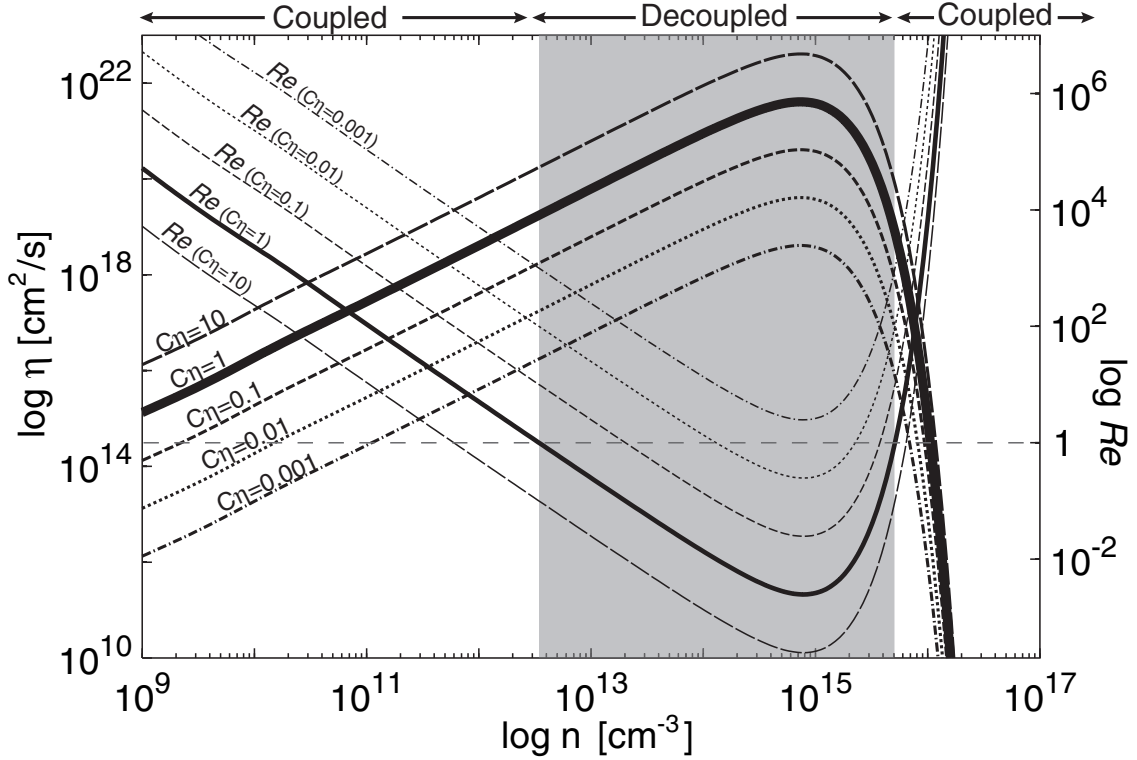}
\caption{
Resistivity $\eta$ (left axis) and magnetic Reynolds number $Re_{\rm m}$ (right axis) as a function of the number density.
The resistivity estimated by \citet{nakano02}  is plotted as a thick solid line ($\ceta=1$) and the corresponding magnetic Reynolds number as a thin solid line.
The magnetic Reynolds number is analytically acquired by using the free-fall velocity $v_{\rm f}$ and Jeans length $\lambda_{\rm j}$ estimated for the given density and resistivity as $Re_{\rm m} \equiv v_{\rm f}\lambda_{\rm j} \eta^{-1}$. 
Resistivities multiplied by factors of 0.001 ($\ceta=0.001$), 0.01 ($\ceta=0.01$), 0.1 ($\ceta=0.1$), and 10 ($\ceta=10$), and the corresponding magnetic Reynolds numbers are also plotted.
The magnetic field is well coupled with the gas in the ``coupled'' region, while the magnetic field is decoupled from the gas in the ``decoupled'' region.
Below the vertical broken line $Re_{\rm m}=1$,  Ohmic dissipation is effective. 
Thus, in the shadowed region ($3\times 10^{12} \lesssim n \lesssim 5 \times 10^{15}\cm$), the magnetic field is effectively dissipated by  Ohmic dissipation for $\ceta=1$.
}
\label{fig:1}
\end{center}
\end{figure}

To describe realistic evolution of magnetic field in the protostar 
 formation, we should take into account the {\em non-ideal} MHD effects 
 of weakly ionized molecular gas.  
The detailed physical processes in the problem at hand are studied by 
 many authors \citep[e.g.,][references therein]{nakano02}.  
In general, the ambipolar diffusion is slow but important 
 in the low density phase, 
 and Ohmic dissipation dominates in the high-density phase.  
In the intermediate density phase, the Hall term effect can be also 
 important depending on the size distribution of dust grains 
 \citep{WardleNg99}. 
Note, however, that 
 Ohmic dissipation is the most efficient mechanism for 
 the dissipation of magnetic field in the magnetically supercritical 
 cloud core, 
 and we are mainly interested in this dynamically contracting gas.  
Therefore we model the dissipation of magnetic field only by 
 the effective resistivity in the induction equation (\ref{eq:reg}). 
We quantitatively estimate the resistivity $\eta$ 
 according to \citet{nakano02} and assume that $\eta$ is a function of density and temperature as 
\begin{equation}
\eta = \dfrac{740}{X_e}\sqrt{\dfrac{T}{10 {\rm K}}} \left[ 1-{\rm tanh}\left( \dfrac{n}{10^{15}\cm}  \right)  \right] \, \, \, {\rm cm}^2\,{\rm s}^{-1},
\label{eq:etadef}
\end{equation}
where $T$ and $n$ are the gas temperature and number density, and $X_e$ is the ionization degree of the gas which describes as
\begin{equation}
X_e =  5.7 \times 10^{-4} \left( \dfrac{n}{{\rm cm}^{-3}} \right)^{-1}.
\end{equation} 
We added the second term in the right-hand side of equation~(\ref{eq:etadef}) to smoothly decline the diffusivity at $n \simeq 10^{15}\cm$, which means Ohmic dissipation becomes ineffective for $n > 10^{15}\cm$.
When the gas density reaches $n \simeq 10^{15}\cm$, the temperature reaches $T \sim 2000$K.
Thus, the thermal ionization of alkali metals reduces the resistivity and  the magnetic field is coupled with the gas $n \gtrsim 10^{15}\cm$.
We are using a barotropic equation of state,  and hence, the temperature is a function of the density.  
Therefore, $\eta$ can be expressed as a function of density,  represented by the thick line in Figure~\ref{fig:1}.
To take into account the uncertainty of the effective resistivity,  we parameterize the maximum value of $\eta$,  keeping the shape of the function, and define $\ceta$ as
\begin{equation}
\eta = \ceta \, \eta_0(\rho),
\end{equation}
where $\eta_0(\rho)$ is a function of the central density and corresponds to the thick line in Figure~\ref{fig:1}.
$\eta$ corresponds to $\eta_0 (\rho)$ when $\ceta=1$.
In this paper, we adopt $\ceta=0$, $10^{-3}$, 0.01, 0.1, 1, and 10,  as listed in Table~\ref{table:init},  to investigate the dependence on $\eta$, and hence, 
 the effect of Ohmic dissipation. 
Since the second term on the right-hand side in Equation~(\ref{eq:reg}) vanishes in models with $\ceta=0$, these models obey the ideal MHD approximation.
In this paper, we call models having $\ceta\ne0$ `non-ideal MHD models' and models having  $\ceta=0$  `ideal MHD models.'

We analytically estimate the magnetic Reynolds number for models with different $\ceta$  ($Re_{\rm m} \equiv v_f \, \lambda_j \, \eta^{-1}$; thin lines in  Fig.~\ref{fig:1}), where $v_f \equiv [(4/3) \pi G \lambda_j^2 \rho_c]^{1/2} $ is the free-fall velocity and $\lambda_j \equiv (\pi c_s^2 / G \rho_c)^{1/2}$ is the Jeans length.
 Magnetic dissipation is effective for $2\times 10^{12} \cm \lesssim n_c \lesssim 6 \times 10^{15} \cm$ in models with $\ceta=1$ (thin solid line), which corresponds to the results of \citet{nakano02}, while it is effective for  $2\times 10^{14} \cm \lesssim n \lesssim 2 \times 10^{15} \cm$ in models with $\ceta=0.01$.
In contrast, at density of $n \sim 10^{15}\cm$ for example, a mean free path of main particles is about 1cm.
Thus, (hydrodynamic) Reynolds number is extremely large,  and magnetic Prandtl number is correspondingly small.
Note that the viscous dissipation is important only at the shock front, and we can safely neglect the effect of (hydrodynamical) viscosity in the calculation with our shock-capturing numerical scheme.

\section{Numerical Method}
  We adopt the nested grid method  \citep[for details, see][]{machida05a,machida06a} to obtain high spatial resolution near the center.
  Each level of a rectangular grid has the same number of cells ($ 64 \times 64 \times 32 $),  although the cell width $h(l)$ depends on the grid level $l$.
 The cell width is reduced by a factor of 1/2 as the grid level increases by 1 ($l \rightarrow l+1$).
 We assume mirror symmetry with respect to $z$ = 0.
 The highest level of a grid changes dynamically.
 We begin our calculations with four grid levels ($l=1$, 2, 3, 4).
 The box size of the initial finest grid $l=4$ is chosen to be $2 R_{\rm c}$, where $R_c$  denotes the radius of the critical Bonnor--Ebert sphere. 
 The coarsest grid, $l=1$, therefore, has a box size of $2^4\, R_{\rm c}$. 
 A boundary condition is imposed at $r=2^4\, R_{\rm c}$, such that the magnetic field and ambient gas rotate at an angular velocity of $\Omega_0$ (for details, see \citealt{matsu04}).
  A new finer grid is generated whenever the minimum local  
$ \lambda _{\rm J} $ becomes smaller than $ 8\, h (l_{\rm max}) $. 
The maximum level of grids is restricted to $l_{\rm max} = 30$.
 Since the density is highest in the finest grid, the generation of a new grid ensures the Jeans condition of \citet{truelove97} with a margin of safety factor of 2.
 We adopted the hyperbolic divergence $\vect{B}$ cleaning method of \citet{dedner02}.

\section{Results}
\begin{figure}
\begin{center}
\includegraphics[width=150mm]{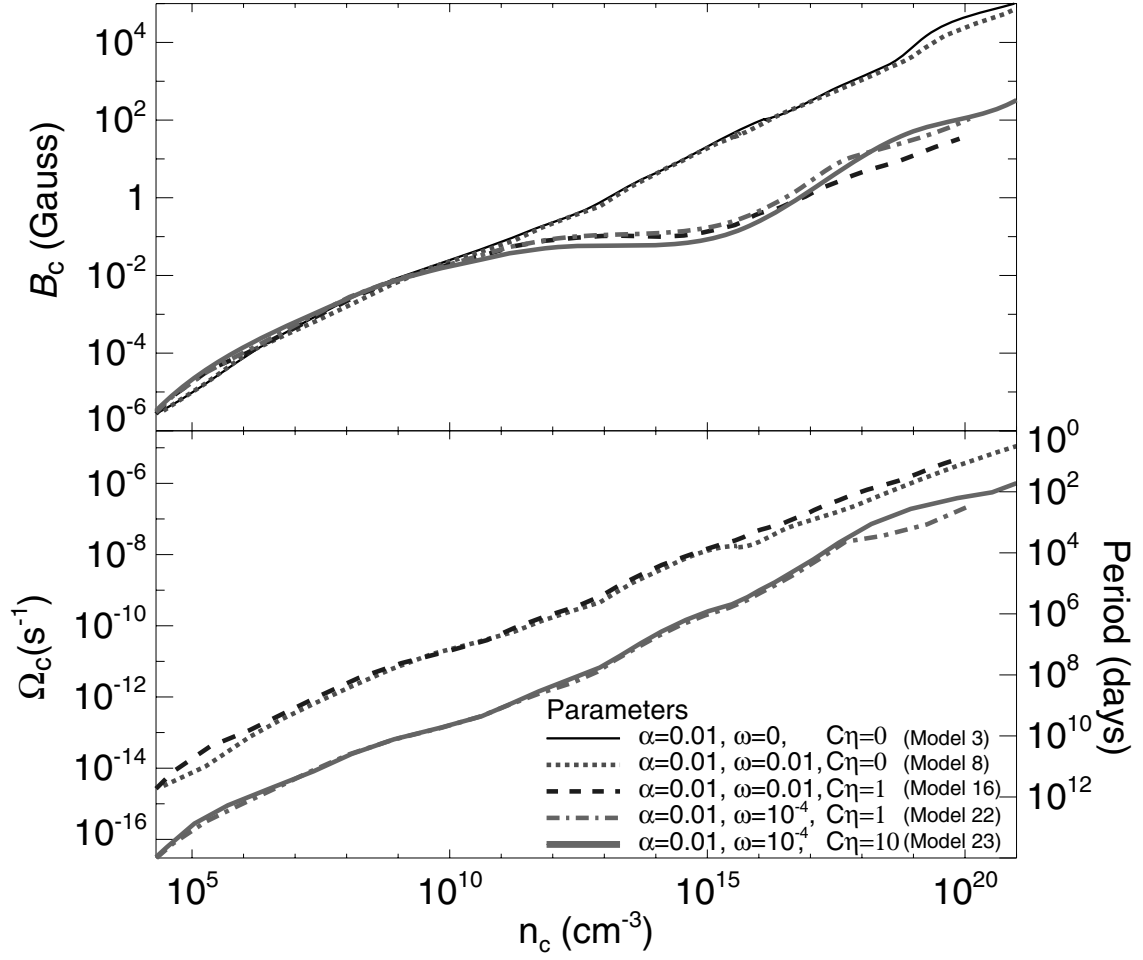}
\caption{
Evolution of the central magnetic field $B_{\rm c}$ (upper panel) and angular velocity $\Omega_{\rm c}$ (lower panel) against the number density at the center of the cloud for Models 3, 8, 16, 22, and 23.
The rotation periods $P=2\pi/\Omega_{\rm c}$ are also plotted on the right-hand axis of the lower panel.
}
\label{fig:2}
\end{center}
\end{figure}

Starting from the number density $\nc = f \times 10^4 \cm$ ($f=1.2$ or 1.4), we calculate the cloud evolution until a protostar is formed ($\nc \simeq 10^{22} \cm$).
We assume that the initial clouds have magnetic fields of $B_0 = 0$--34\,$\mu$G, and angular velocities of $\Omega_0=0$--3.1$\times 10^{-14}$\,s$^{-1}$, as listed in Table~\ref{table:init}.
These values are comparable with observed magnetic fields  \citep{crutcher99} and angular velocities \citep{goodman93, caselli02} of molecular cloud cores.
Figure~\ref{fig:2} shows the evolution of the magnetic field $B_{\rm c}$ and angular velocity $\Omega_{c}$ at the center of clouds for Models 3, 8, 16, 22, and 23.
The growth rates of the magnetic field for the non-ideal MHD Models 16, 22, and 23 ($\ceta \ne 0$) are smaller than those for the ideal MHD Models 3 and 8 ($ \ceta \ne 0$) for $10^{11} \cm \lesssim \nc \lesssim 10^{16}\cm$ (Fig.~\ref{fig:2} upper panel).
This is because the magnetic field is effectively dissipated by  Ohmic dissipation for this phase.
The magnetic fields at the star formation epoch ($\nc \simeq 10^{21}\cm$) $B_{\rm f}$ are listed in Table~\ref{table:init}.
Models with $\ceta=0$ (ideal MHD models) have magnetic fields of $\sim 100 \kg$ (Model 3: 106$\kg$, Model 8: 67.3$\kg$), while models with $\eta \ne 0$ (non-ideal MHD models) have magnetic fields of $0.1$--$1\kg$ (Model 16: 0.11$\kg$, Model 22: 0.68$\kg$, Model 23: 0.32$\kg$).
The magnetic field strengths of protostars observed from Zeeman broadening measurements are $\sim 1\kg$ \citep{johns99a,johns99b, johns01,bouvier06}.   
Even though protostars in non-ideal MHD models have masses of only $\sim 10^{-3}\msun$ in our simulation, they have magnetic field strengths equivalent to those of observed protostars.
The magnetic field is considered to be amplified by magneto-rotational instability (MRI; \citealt{balbus91}) and convection on the surface of a protostar in the later phase of star formation.
However, our results indicate that protostars already have magnetic fields of $\sim$$\kg$ for non-ideal MHD models at their formation epoch.
On the other hand, when a protostar is formed, the magnetic field for ideal MHD models reaches $B\sim$100$\kg$, which is about 100 times larger than the magnetic field of observed protostars.

The rotation periods at the star formation epoch ($\nc \simeq 10^{21}\cm$) $P$ are listed in Table~\ref{table:init}.
Figure~\ref{fig:2} shows the evolutions of the angular velocity $\Omega_{\rm c}$ (left axis) and rotation period $P$ (right axis) at the centers of clouds.
This figure shows that the rotation period reaches $P$ = 1--100 days at the protostar formation epoch ($\nc = 10^{21}\cm$), even when the initial cloud has a small angular velocity ($2\times 10^{-17}$s$^{-1}$).
Many observations indicate that protostars have rotation periods of $P=0.6$--20 days \citep[e.g.,][]{herbst06}. 
Thus, the rotation periods derived from our calculation are roughly in agreement with those of observed protostars.
The angular momentum is considered to be modified by interactions between the protostar and protoplanetary disk, jets, and outflow in the later phase of star formation.
However, in our calculations, protostars have rotation rates of the same order of magnitude at their formation epoch, as is the case for the magnetic field.

We show the evolution of the magnetic field in a collapsing cloud in the following sections in detail.
Firstly, we show the cloud evolution for ideal MHD models ($\ceta=0$) to simply investigate the evolution of the magnetic field.
The cloud evolution for non-ideal MHD models ($\ceta \ne 0$) is shown in subsequent sections.

\subsection{Evolution of the Magnetic Field for Ideal MHD Models}
\subsubsection{Cloud Evolution with Different Parameters}
\begin{figure}
\begin{center}
\includegraphics[width=150mm]{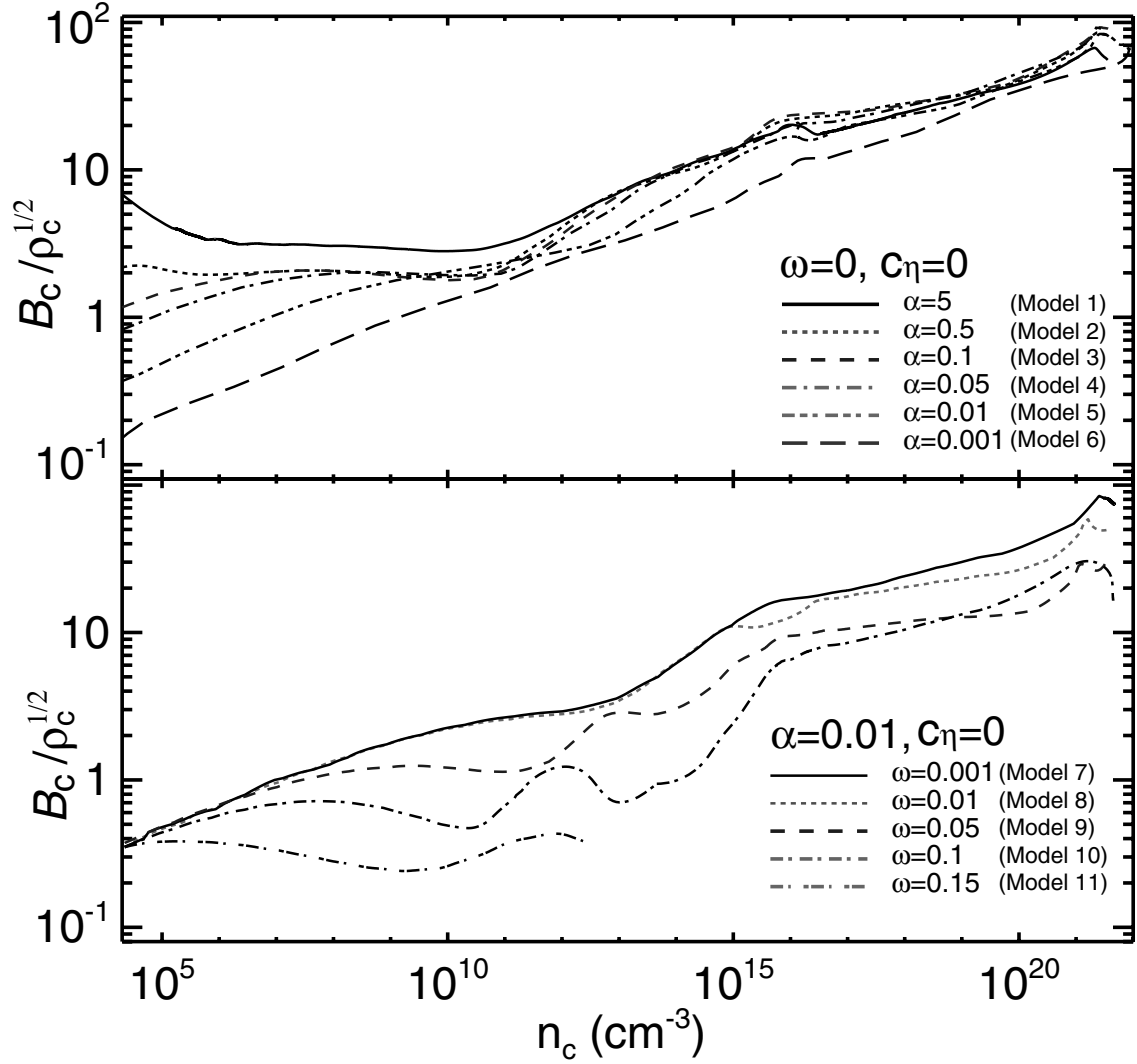}
\caption{
Magnetic field $B_{\rm c}$ normalized by the square root of the density $\rhoc^{1/2}$ in the unit of initial sound speed $c_{\rm s,0}$ plotted against the number density at the center of the cloud for Models 1--6 (upper panel) and 7--11 (lower panel).
}
\label{fig:3}
\end{center}
\end{figure}

Figure~\ref{fig:3} shows the evolution of the magnetic field $B_{\rm c}$ normalized by the square root of the density $\rhoc^{1/2}$ in the unit of initial sound speed $c_{s,0}$ at the centers of clouds for ideal MHD models ($\eta=0$).
As shown in \citet{machida05a, machida06a}, the evolution of $B_{\rm c}/\rhoc^{1/2}$ depends on the mode of the cloud collapse, spherical, vertical, or disk-like collapse.
When the cloud has a small amount of rotational or magnetic energy, the cloud collapses almost spherically (spherical collapse).
On the other hand, the cloud collapses vertically along the rotation axis or magnetic field lines (vertical collapse) when the cloud has a large amount of rotational or magnetic energy.
In either case, a thin disk finally forms around the center of a cloud, and the cloud continues to collapse keeping a disk-like structure (disk-like collapse).
We briefly summarize the relation between the growth rate of the central magnetic field and the cloud density \citep[for details see][]{machida05a,machida06a}.
The magnetic field strength increases as $B_{\rm c}\propto \rhoc^{2/3} $ in weakly magnetized and slowly rotating clouds (spherical collapse), because the cloud evolution is mainly controlled by thermal pressure and gravity. 
The magnetic field remains constant at $B_{\rm c}\propto \rhoc^0$ when the cloud is magnetized strongly or rotating rapidly (vertical collapse), because the radial contraction of the cloud is suppressed by the strong magnetic tension or centrifugal force.
After the disk formation, the magnetic field evolves as $B_{\rm c} \propto \rhoc^{1/2}$, and the Lorentz and centrifugal forces are balanced with the thermal pressure gradient and gravitational forces in the collapsing cloud (disk-like collapse)\footnote{
In this paper, we use the terminology `disk-like collapse' when the disk is formed and the magnetic field increases as $B_{\rm c}\propto\rhoc^{1/2}$ at the center of the cloud.
}.
For the isothermal phase ($\nc \lesssim 10^{11}\cm$), $B/\rho^{1/2}$ at the center of a cloud converges to a certain value (the magnetic flux--spin relation; \citealt{machida05a}) when the cloud has a smaller rotational energy than the magnetic energy at the initial state.
Thus, the initial strength of the magnetic field is not sensitive to the cloud evolution, especially for  $\nc > 10^{11}\cm$, if the field strength is sufficiently strong.
Observations indicate that the magnetic energy is much larger than the rotational energy in molecular cloud cores \citep{crutcher99,caselli02}.
Thus, convergence of $B/\rho^{1/2}$ is expected in the collapsing region of real molecular clouds.

Figure~\ref{fig:3} upper panel shows evolutions of the magnetic field in non-rotating clouds ($\omega =0$) having different magnetic fields at the initial stage (Models 1, 2, 3, 4, 5, and 6).
The magnetic field of Model 1 ($B_0 = 34\,\mu$G) is about 70 times stronger than that of Model 6 ($B_0 = 0.5\,\mu$G) at the initial stage.
The figure shows that the magnetic fields for all the models converge to certain values [$B_{\rm c}/\rhoc^{1/2} \simeq 0.55$ (Model 6) to 1.2 (Model 12) ] at the end of the isothermal phase ($\nc \simeq 10^{11}\cm$).
$B_{\rm c}/\rhoc^{1/2}$ begins to increase after the gas becomes adiabatic  ($\nc \gtrsim 10^{11}\cm$), because the geometry of the collapse changes from disk-like to spherical for increasing thermal pressure.
However, the clouds have almost the same magnetic field strengths for this epoch ($\nc \gtrsim 10^{11}\cm$), because the growth rates of the magnetic field are almost the same for these models.
The formed protostars have magnetic fields of $B_{\rm c} = 88.9$--$123\kg$ at the protostar formation epoch ($\nc \simeq 10^{21}\cm$), as listed in Table~\ref{table:init}.
Thus, the formed protostars have magnetic field strengths different by a factor of 1.4, while clouds strength differ by a factor of 70 at the initial stage.
As a result, protostars formed from initially non-rotating clouds have almost the same magnetic field if Ohmic dissipation is ignored.
As shown in later sections, since both cloud rotation and Ohmic dissipation only decrease the magnetic field strength of a formed protostar, a protostar at its formation epoch cannot have a magnetic field exceeding $B_{\rm c} \simeq 100 \kg$, derived from the non-rotating and ideal MHD models.

The growth rate of the magnetic field for clouds with the same initial magnetic field depends on the cloud rotation, because the geometry of the collapse depends not only on the magnetic field but also on the cloud rotation \citep{machida05a,machida06a}.
Figure~\ref{fig:3} lower panel shows the evolution of the magnetic field strength normalized by the square root of the density $B_{\rm c}/\rhoc^{1/2}$ for ideal MHD models ($\ceta=0$) with different initial angular velocities and the same initial magnetic field strengths ($B_{\rm c}= 1.6\,\mu$G).
The figure indicates that the growth rate of the magnetic field is smaller for an initially rapidly rotating cloud.
Note that the calculation was stopped at $\nc \simeq 10^{12}\cm$ in Model 11, with the largest angular velocity at the initial state, because fragmentation occurred.
At the protostar formation epoch ($\nc = 10^{21}\cm$), the cloud with the initially slowest angular velocity ($\omega = 0.001$; Model 7) had a magnetic field of $B_{\rm c} = 95.9\kg$, while the rapidly rotating cloud with $\omega=0.1$ (Model 10) had $B_{\rm c}=43.7\kg$ (see Table~\ref{table:init}).
This is because the growth rate of the magnetic field changes from $B\propto \rho^{2/3}$ to $B\propto \rho^{1/2}$ at an earlier evolutional stage in more rapidly rotating clouds.
Therefore, the magnetic field at the protostar formation epoch depends on the rotation of the initial cloud.
However, a slowly rotating cloud has a magnetic field strength only twice that of a rapidly rotating cloud (see Models 7--11 in Table~\ref{table:init}).

\subsubsection{Cloud Evolution of Typical Model}
\begin{figure}
\begin{center}
\includegraphics[width=130mm]{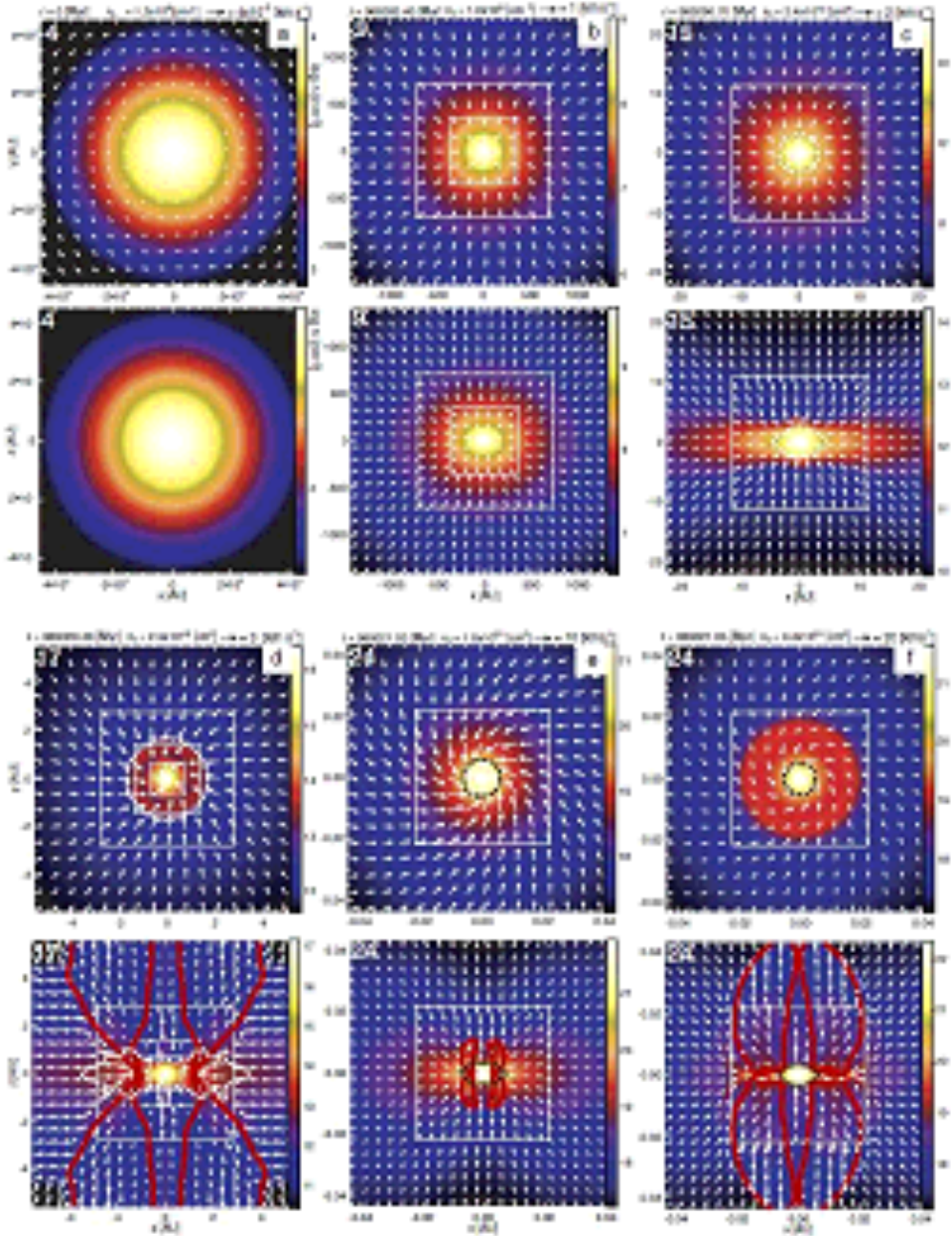}
\vspace{-0.5cm}
\caption{
Density (color-scale) and velocity distribution (arrows) on the cross-section in the $z=0$ plane (upper panels) and $y=0$ plane (lower panels) for Model 12 [($\alpha, \omega, \ceta$) = (0.01, 0.01, 0)].
Panels (a) through (f) are snapshots at the stages 
(a) $n_c =1.2 \times 10^4 \cm$ ($l=4$; initial state), \ 
(b) $ 1.0 \times 10^9 \cm$ ($l=9$--11; isothermal phase), \ 
(c) $ 3.4 \times 10^{13} \cm$ ($l=15$, 16; adiabatic phase), \
(d) $ 2.5 \times 10^{16} \cm$ ($l=17$--20; second collapse phase), \
(e) $ 1.8 \times 10^{21} \cm$ ($l=24$, 25; protostellar phase), \ and \
(f) $ 4.4 \times 10^{21} \cm$ ($l=24$, 25; calculation end),
where $l$ denotes the level of the subgrid.
The dotted lines indicate the first core (white) and second core (black), surrounded by the shock surfaces. 
The red thick lines indicate the border between the infalling and outflowing gas ($v_{\rm out} \ge 0$ km s$^{-1}$).
The level of the subgrid is shown in the upper left corner of each upper panel.
The elapsed time $t$, density at the center $\nc$, and arrow scale are denoted in each panel.
}
\label{fig:4}
\end{center}
\end{figure}

\begin{figure}
\begin{center}
\includegraphics[width=160mm]{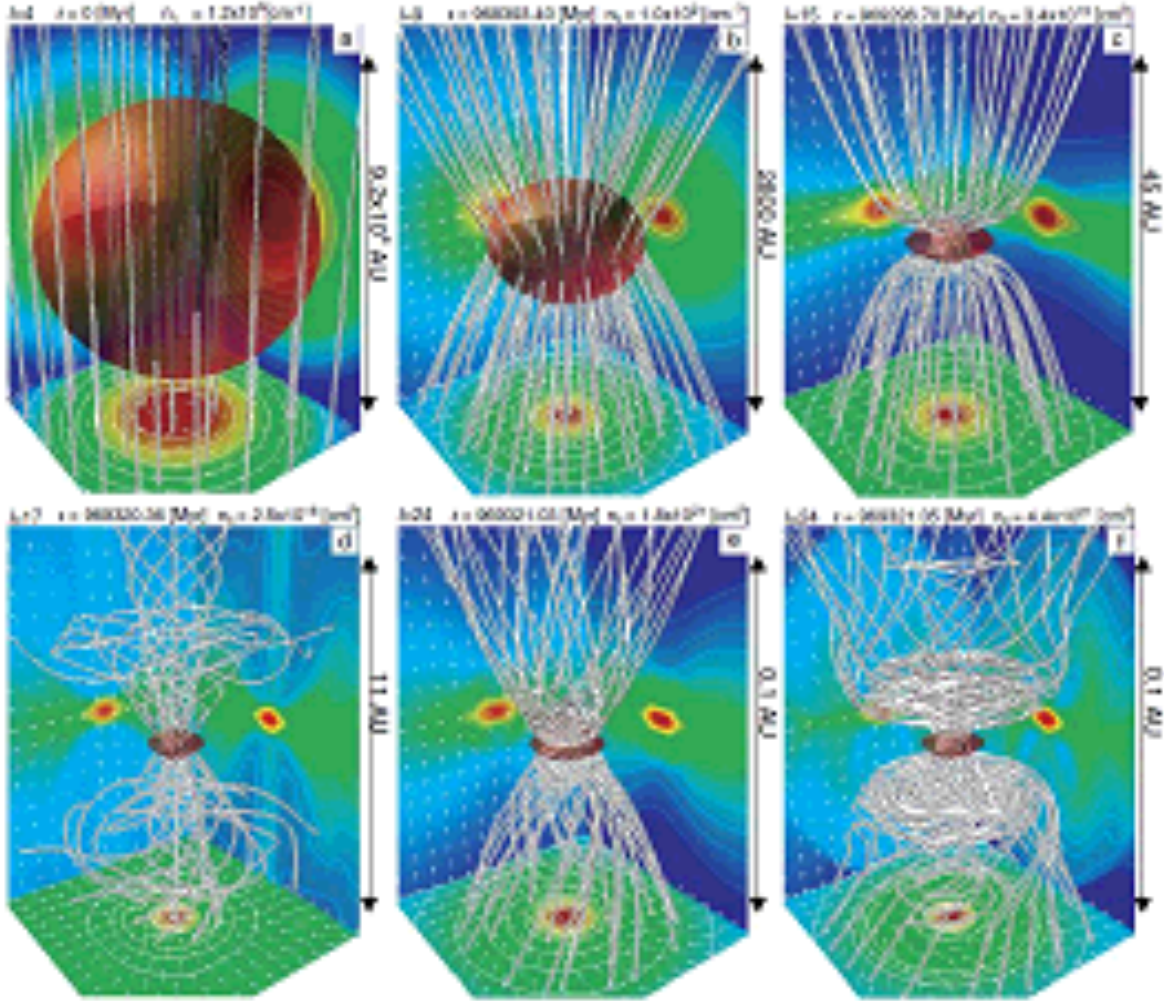}
\caption{
Three-dimensional projections of Fig.~\ref{fig:4}.
The panels (a)--(f) have the same scale and epoch as panels (a)--(f) of Fig.~\ref{fig:4}.
The structure of the high-density region ($\rho > 0.1\rho_{\rm c}$; red iso-density surface)  and magnetic field lines (black-and-white streamlines) are plotted in each panel.
The density contours (false color and contour lines), velocity vectors (arrows) on the cross-section in the $x=0$, $y=0$, and $z=$0 plane are, respectively, projected on the sidewalls of the graphs.
The grid level $l$, elapsed time $t$, density at the center $\nc$,  and grid scale are denoted in each panel.
}
\label{fig:5}
\end{center}
\end{figure}

Figure~\ref{fig:4} shows the cloud evolution from the initial state ($\nc = 1.4 \times 10^{4}\cm$; Fig.~\ref{fig:4}{\it a}) to the protostar formation epoch ($\nc = 4.4\times 10^{21}\cm$; Fig.~\ref{fig:4}) for Model 12 with parameters $\alpha=0.01$, $\omega=0.01$, and $\ceta=0$.
In this model, the magnetic field is well coupled to the gas from the initial to the final stage because $\ceta=0$.
The initial cloud is weakly magnetized ($B_0 = 1.4\,\mu$G) and rotating slowly ($\Omega_0 = 2\times 10^{-15}\,$s$^{-1}$).
This cloud is classified as a magnetic-force dominant model by the criterion in \citet{machida05a,machida06a}, because the ratio of the initial angular velocity $\Omega_{\rm 0}$ to the magnetic field $B_0$ is $\Omega_0/B_0 = 1.4\times 10^{-9} < 5.3 \times 10^{-9}\equiv (\Omega/B)_{\rm cri}$.
Thus, the magnetic field affects the cloud evolution, while the effect of cloud rotation is small (for details, see \citealt{machida05a,machida06a}). 
Figure~\ref{fig:5} shows the magnetic field distribution at the same epoch for the panels in Figure~\ref{fig:4}.
As shown in Figure~\ref{fig:4}{\it a} and {\it b}, the central region gradually becomes oblate as the cloud collapses and a thin disk is formed in the isothermal phase ($\nc \lesssim 10^{11} \cm$).
The magnetic field lines gradually converge towards the center (Fig.~\ref{fig:5}{\it b} and {\it c}).
This configuration of the magnetic field lines is similar to that observed by \citet{girart06}.
The thin solid lines in Figure~\ref{fig:6} left panels show the evolution of the magnetic field normalized by the square root of the density ($B_{\rm c}/\rhoc^{1/2}$; Fig.~\ref{fig:6}{\it a}) in the unit of initial sound speed $c_{s,0}$,  plasma beta ($\betap \equiv 8 \pi c_s^2 \rho/B^2$; Fig.~\ref{fig:6}{\it b}) within the region $\rho > 0.1 \rho_c$, and specific angular momentum normalized by the mass ($J/M^2$; Fig.~\ref{fig:6}{\it c}) within the region $\rho > 0.1 \rho_c$ in the unit of $4\pi G/c_{s,0}$, against the central density.
Figure~\ref{fig:6}{\it a} and {\it b} show that both $B_{\rm c}/\rhoc^{1/2}$ and $J/M^2$ continue to increase in the isothermal phase ($\nc \lesssim 10^{11}\cm$).
Thus, it is considered that the disk is formed by the Lorentz and centrifugal forces, as shown in Figure~\ref{fig:5}{\it c}.
The plasma beta has a minimum of $\betap \simeq 8$ at the end of the isothermal phase ($\nc \simeq 10^{11}\cm-10^{12}\cm$; Fig.~\ref{fig:6} middle panel).
Even after the gas around the center of the cloud becomes adiabatic ($\nc \gtrsim 10^{11}\cm$), the magnetic field continues to be amplified (Fig.~\ref{fig:6}{\it a}).
On the other hand, the plasma beta remains $\betap \simeq 10$--20 for $\nc \gtrsim 10^{17}\cm$, after it increases slightly owing to an increase of the thermal energy for $10^{12}\cm \lesssim \nc \lesssim 10^{16}\cm$.

When the gas reaches $\nc\simeq 10^{11}\cm$, the central region becomes optically thick and the equation of state becomes hard, as derived from the one-dimensional radiative hydrodynamical calculation \citep{masunaga00}.
After the equation of state becomes hard, shock occurs and the first core is formed.
The first core is formed at $\nc \sim 10^{13}\cm$ in this model.
The first core is shown by the white-dotted line in Figure~\ref{fig:4}{\it c} and {\it d}. 
(We plot the position where the shock occurred on the $z=0$ plane in these panels.)
The first core has a mass of $\sim 0.012 \msun$ and radius $\sim 7$\,AU.
The arrows in Figure~\ref{fig:4}{\it d} indicate that the radial component of the velocity $v_{\rm r}$ is dominant outside the first core, while the azimuthal component of the velocity $v_{\phi}$ is comparable to the radial component inside the first core on the $z=0$ plane.
This is because the large thermal pressure suppresses the cloud collapse inside the first core, and thus the rotation timescale becomes smaller than the collapse timescale.
Figure~\ref{fig:4}{\it d} lower panel shows that the outflow is driven from the first core.\footnote
{
We call the flow driven from the first core  `outflow' and the flow driven from the second core a `jet'.
}
Outflow appears when the central density reaches $\nc \simeq 10^{15}\cm$.
Outflow driven from the first core has also been shown by \citet{tomisaka02}, \citet{matsu04}, \citet{machida05b}, \citet{banerjee06}, and \citet{fromang06}.
As shown in Figure~\ref{fig:5}{\it d}, the magnetic field lines begin to twist due to the rotation of the first core.
This outflow reaches $\sim10$\,AU and has a maximum speed of $\sim3\km$ at the end of the calculation.

The specific angular momentum normalized by the mass $J/M^2$ begins to decrease in the adiabatic phase.
\citet{matsu97} and \citet{saigo06} show that $J/M^2$ is constant after it reaches a peak in unmagnetized clouds.
Thus, the decrease of $J/M^2$ is caused by magnetic effects.
Since no outflow appears before the first core formation, the decrease of $J/M^2$ for this phase is caused by magnetic braking \citep{basu94}.
After the central density reaches $\nc \simeq 10^{15}\cm$, $J/M^2$ decreases more rapidly.
This shows that the angular momentum is removed not  only by magnetic braking, but also by outflow in the adiabatic phase.
\citet{tomisaka98} and \citet{matsu04} show that outflow is an important mechanism for angular momentum transfer.

The cloud collapses again inside the first core for $10^{16}\cm \lesssim \nc \lesssim 10^{21}\cm$, because of dissociation of molecular hydrogen (i.e., the second collapse).
Then, the equation of state becomes hard again and the second core (i.e., a protostar) is formed at $\nc \simeq 10^{21}\cm$ \citep{masunaga00}.
Figure~\ref{fig:4}{\it e} shows the structure near the protostar (the protostar or shocked region is represented by the black-dotted line).
The gas around the protostar has a density range of $10^{17} \cm \lesssim \nc \lesssim 10^{20} \cm$ and collapses rapidly, because the equation of state is soft in this region [see Eq.~(\ref{eq:eos})].
Thus, the gas accretes onto the protostar with a high speed of $\sim10\km$.
Since the rotation velocity is comparable to the accretion velocity, the magnetic field lines are strongly twisted (Fig.~\ref{fig:5}{\it d}).
Figure~\ref{fig:4}{\it f} shows the structure around the protostar 215 hours after the protostar formation epoch.
The protostar has a mass of $2.1\times 10^{-3}\msun$ and a radius of $1.1\rsun$ at this epoch.
The disk surrounding the protostar extends up to $3.9\rsun$ with $1.2\times 10^{-4}\msun$.
A strong jet is driven from the protostar, as shown in the upper panel of Figure~\ref{fig:4}{\it f}.
The jet driven from the protostar is also shown by \citet{tomisaka02}, \citet{banerjee06}, and \citet{machida06b}.
The jet reaches $11\rsun$ at the end of the calculation.
In this model, since the effect of Ohmic dissipation is ignored ($\ceta=0$), the magnetic field couples with the gas at any time.
Thus, the magnetic field around the protostar is strongly twisted at the protostar formation epoch, as shown in Figure~\ref{fig:5}{\it f}.
The cloud evolution of Model 12 is similar to that of \citet{tomisaka02} and \citet{banerjee06}, in which the evolution of a magnetized cloud in an ideal MHD regime is studied.

\subsection{Evolution of Magnetic Field in Non-Ideal MHD Models}
\begin{figure}
\begin{center}
\includegraphics[width=140mm]{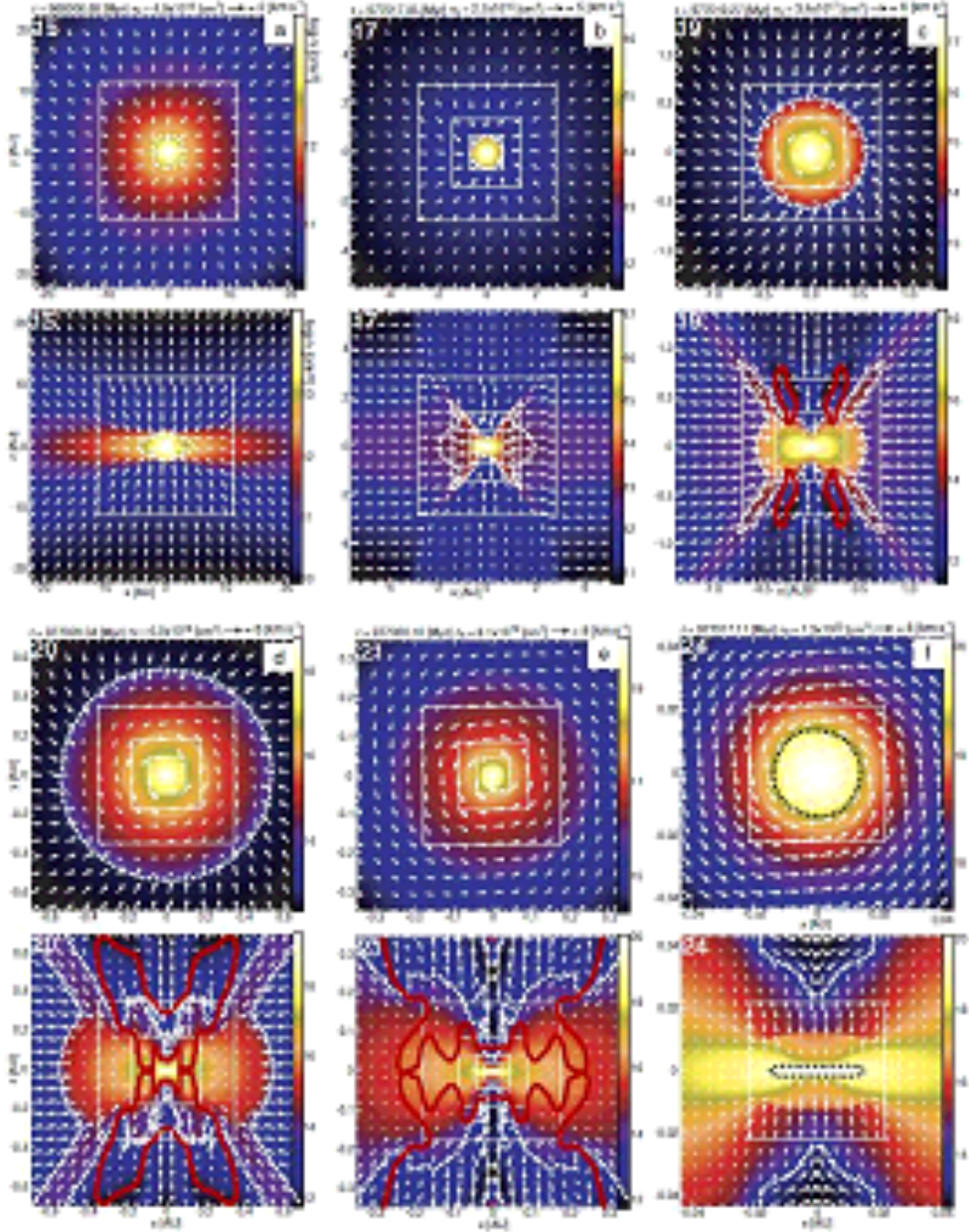}
\vspace{-0.3cm}
\caption{
Same as  Fig.~\ref{fig:4} but for Model 16.
Panels (a) through (f) are snapshots at the stages 
(a) $n_c =4.9 \times 10^{13} \cm$ ($l=15$, 16), \ 
(b) $ 2.7 \times 10^{16} \cm$ ($l=17$--20), \ 
(c) $ 3.4 \times 10^{17} \cm$ ($l=19$--21), \
(d) $ 6.0 \times 10^{18} \cm$ ($l=20$--23), \
(e) $ 8.1 \times 10^{19} \cm$ ($l=21$--25), \ and \
(f) $ 1.3 \times 10^{20} \cm$ ($l=24$, 25),
where $l$ denotes the level of the subgrid.
The dotted lines indicate the first core (white) and second core (black), surrounded by the shock surfaces. 
The red thick lines indicate the border between the infalling and outflowing gas ($v_{\rm out} \ge 0$ km s$^{-1}$).
The level of the subgrid is shown in the upper left corner of each upper panel.
The elapsed time $t$, density at the center $\nc$, and arrow scale are denoted in each panel.
}
\label{fig:7}
\end{center}
\end{figure}

\begin{figure}
\begin{center}
\includegraphics[width=160mm]{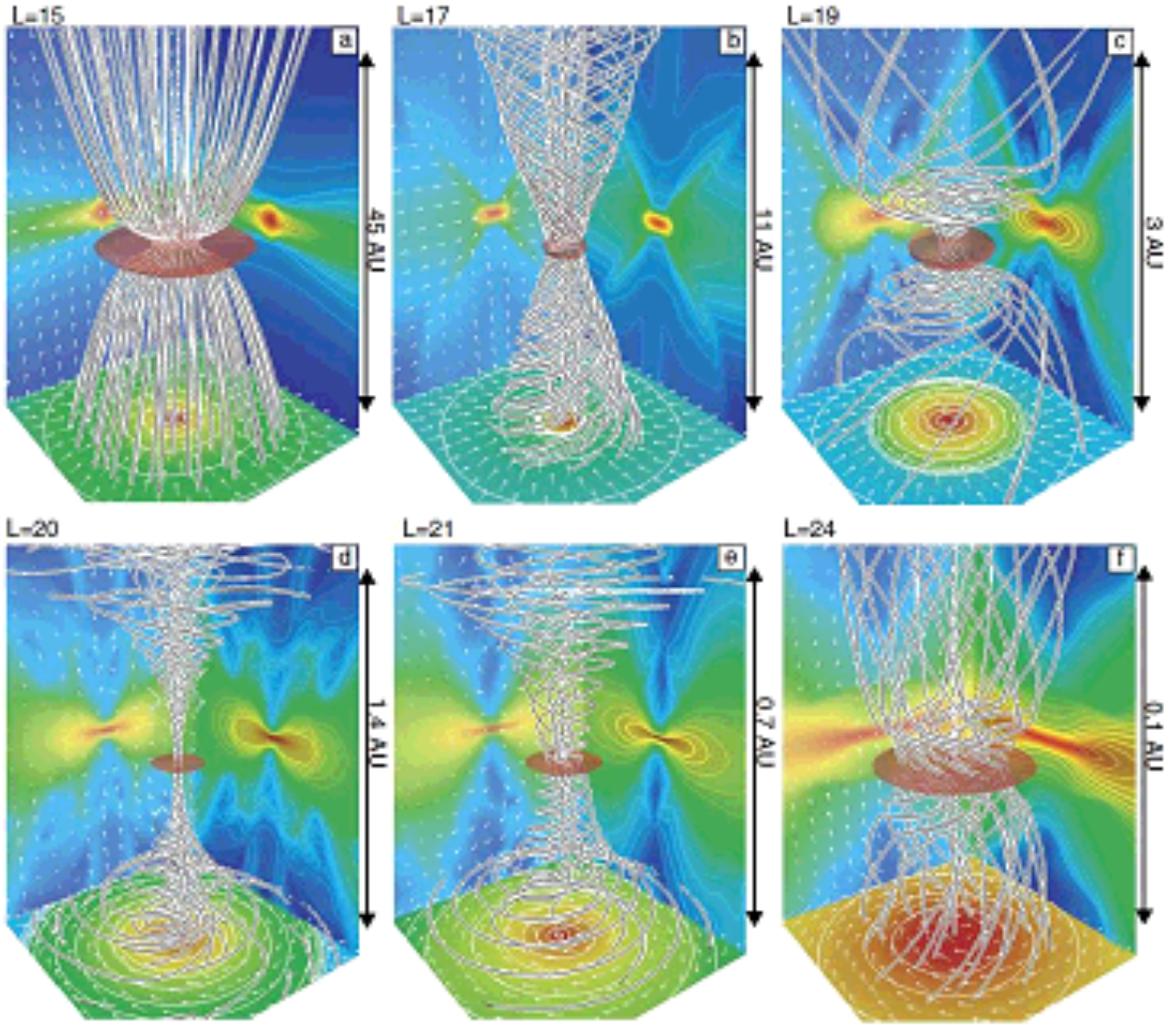}
\caption{
Same as Fig.~\ref{fig:5} but for Model 16.
Panels (a)--(f) are for the same scale and epoch as panels (a)--(f) of Fig.~\ref{fig:7}.
}
\label{fig:8}
\end{center}
\end{figure}

Figures~\ref{fig:7} and \ref{fig:8} show the cloud evolution for Model 16.
Model 16 has the same magnetic field strength ($\alpha=0.01$, $B_0 = 1.4\,\mu$G) and angular velocity ($\omega=0.01$, $\Omega_0 = 1.4\times 10^{-14}$s$^{-1}$) as Model 12 (Figs.~\ref{fig:4} and \ref{fig:5}) at the initial state.
However, Model 16 has $\ceta=1$, which means that the magnetic field is dissipated by Ohmic dissipation in a collapsing cloud (non-ideal MHD model), while the effect of Ohmic dissipation is ignored in Model 12 ($\ceta=0$; ideal MHD model).
In Model 16, the magnetic Reynolds number is $Re_{\rm m} < 1$ for $2\times 10^{12}\cm \lesssim \nc \lesssim 6\times 10^{15}\cm$ (Fig.~\ref{fig:1}).
Thus, the cloud evolutions of Models 12 and 16 are almost the same for $\nc \ll 10^{12}\cm$, because Ohmic dissipation is not effective for this early phase even in Model 16.
The magnetic field normalized by the square root of the density $B/\rho^{1/2}$, plasma beta $\betap$, and specific angular momentum normalized by mass $J/M^2$ in Model 16 (thick solid lines in Fig.~\ref{fig:6}{\it a}--{\it c}) are identical to those in Model 12 (thin solid lines in Fig.~\ref{fig:6}{\it a}--{\it c}) for $\nc \lesssim 10^{11}\cm$.
Thus, Figures~\ref{fig:7} and \ref{fig:8} show the cloud evolution for Model 16 only after the first core is formed.

The first core is formed at $\nc \simeq 10^{13}\cm$ for Model 16, as for Model 12.
Figures~\ref{fig:7}{\it a} and \ref{fig:4}{\it c} show that the first cores in Models 12 and 16 have almost the same shape and size.
The first core has a mass of $0.012\msun$ and a radius of $7$\,AU in Model 16.
As shown in Figure~\ref{fig:8}{\it a},  the poloidal component of the magnetic field is dominant at the first core formation epoch ($\nc \simeq 10^{13}\cm$).
After the formation of the first core, the magnetic field lines begin to be twisted (Fig.~\ref{fig:8}{\it b}).

The outflow from the first core is shown in the lower panel of Figure~\ref{fig:7}{\it c} (red solid lines).
Outflow appears 22\,yr after the first core formation epoch in Model 16 ($\nc \simeq 10^{17}\cm$), while outflow appears 2.6\,yr after the first core formation epoch ($\nc \simeq 10^{15}\cm$) in Model 12.
In Model 16, the outflow has a maximum speed of $\sim2\km$ and reaches $\sim1$\,AU at the end of the calculation.
Figures~\ref{fig:7}{\it c} and \ref{fig:4}{\it d} show that the outflow in Model 16 is smaller and slower than that in Model 12.
The outflow is mainly driven by the Lorentz force outside the first core.
The magnetic field is weak and barely twisted around the first core in Model 16, because the magnetic field in this region is dissipated by Ohmic dissipation.
Thus, outflow in non-ideal MHD models (Model 16) is weaker than that in ideal MHD models (model 12).
We will further discuss the driving mechanism of outflow and jets in a companion paper.

Figure~\ref{fig:6}{\it a} shows that the magnetic field in Model 16 (thick solid line) is largely removed from the central region for $10^{12}\cm \lesssim \nc \lesssim 10^{16}\cm$.
The plasma beta in Model 16 increases from $\betap\simeq10$ to $\simeq 2\times 10^5$ for the same epoch (Fig.~\ref{fig:6}{\it b}).
In Figure~\ref{fig:8}{\it c}, the magnetic field lines are strongly twisted near the center of the cloud, while they are barely twisted away from the center of the cloud.
This implies that the neutral gas is well coupled with the magnetic field, and the magnetic field lines rotate with the neutral gas near the center because Ohmic dissipation is ineffective.
When the gas density reaches $\nc\simeq 10^{16}\cm$, the temperature reaches $T\sim 2000$K.
Thus, the thermal ionization of alkali metals reduces the resistivity and Ohmic dissipation becomes ineffective.
As a result, the magnetic field is coupled with the gas and amplified again for $\nc \gtrsim 10^{16}\cm$.
On the other hand,  the neutral gas is decoupled from the magnetic field in the region away from the center and the magnetic field lines slip through the neutral gas and rotate freely because this region has a density range of $10^{11}\cm \lesssim \nc \lesssim 10^{15}\cm$ (Fig.~\ref{fig:7}{\it c} lower panel) and magnetic dissipation is effective.
Thus, the magnetic field lines that are strongly twisted in the earlier phase ($\nc \ll 10^{12}\cm$) are de-collimated and relaxed by the magnetic tension force.
Note that the magnetic field lines do not disappear, even with Ohmic dissipation. 
In our settings, it is assumed the magnetic field has a vertical component $B_{\rm z}$ in the initial state.
Thus, even when both the azimuthal $B_{\phi}$ and radial $B_{\rm r}$ components of the magnetic field are completely dissipated by Ohmic dissipation, the vertical component of the magnetic field remains because of the condition $\vect{\nabla}\cdot \vect{B}=0$.

As shown in the upper panel of Figure~\ref{fig:7}{\it d}, the gas inside the first core rotates rapidly.
The specific angular momentum in Model 16 is larger than that in Model 12 (Fig.~\ref{fig:6}{\it c}) for $\nc \gtrsim 10^{12} \cm$, while the magnetic field in Model 16 is smaller than that in Model 12 (Fig.~\ref{fig:6}{\it a}).
This means that magnetic braking is less effective in Model 16.
Therefore, a strong centrifugal force creates a thin disk at the center of the cloud, as shown by the projected density contour on the sidewall in Figure~\ref{fig:8}{\it d} and {\it e}.
In addition, the cloud collapses slowly for a rapid rotation, as discussed in \citet{saigo06}.
It takes 40.7\,yr to form the second core ($\nc \simeq 10^{21}\cm$) after the first core formation epoch ($\nc \simeq 10^{13}\cm$) in Model 16, while it takes 25.5\,yr in Model 12. 
The magnetic field lines become twisted again inside the first core (Fig.~\ref{fig:8}{\it d} and {\it e}).
Then, a different flow from that driven from the first core appears, as shown in the lower panels in Figures~\ref{fig:7}{\it d} and {\it e}.
Inside the red line in Figure~\ref{fig:7}{\it d} lower panel, slow flow (outflow) driven from the first core is seen in the region $\vert z \vert > 0.4$, while fast flow (a jet) driven from a rotational supported core inside the first core is seen in the region $\vert z \vert < 0.4$.

\begin{figure}
\begin{center}
\includegraphics[width=160mm]{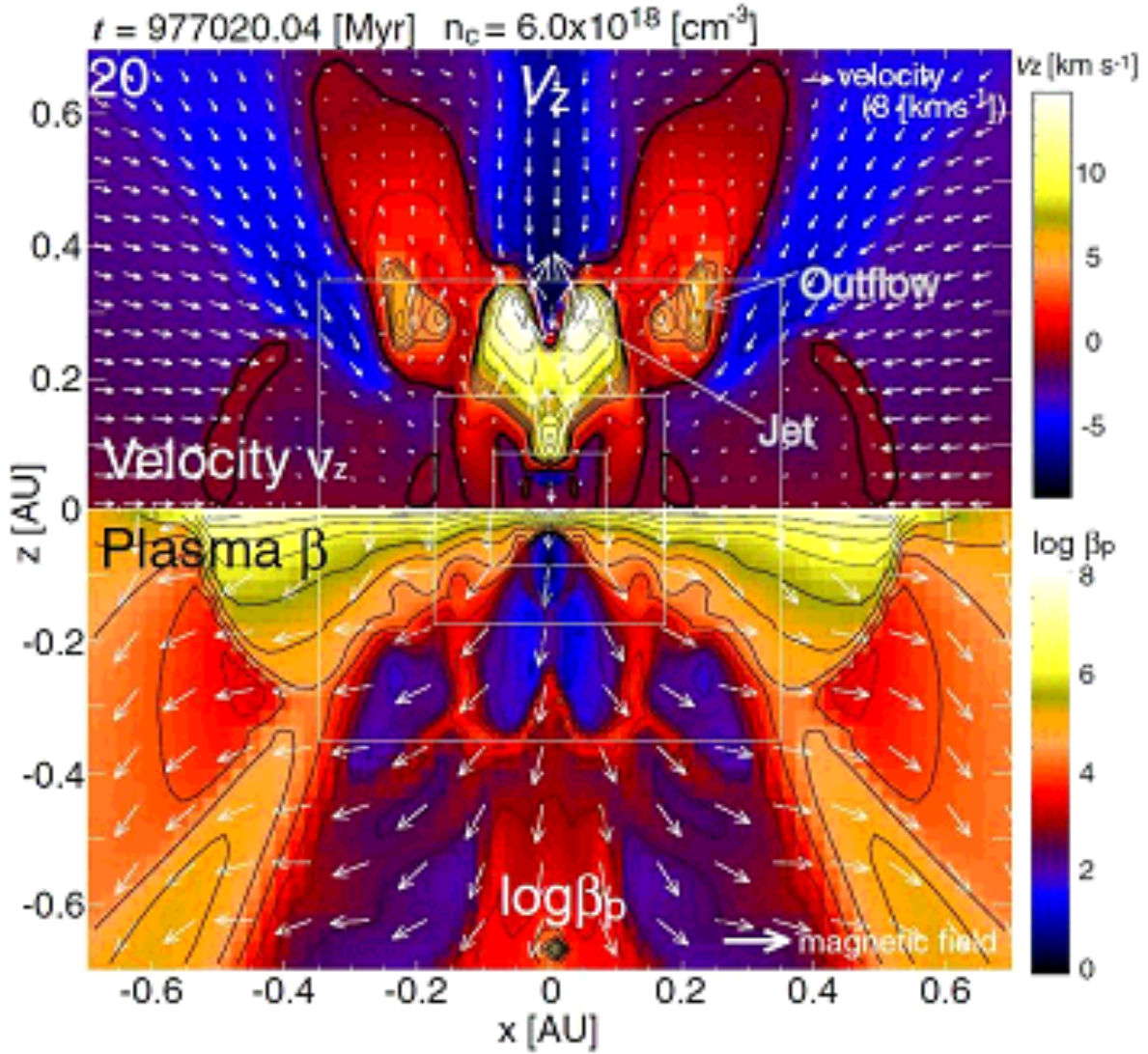}
\caption{
Velocity and plasma beta for Model 16 at the same scale and epoch as Fig.~\ref{fig:7}{\it d}.
Upper panel ($z>0$): Velocity of the $z$-component $v_z$ (color-scale, contours) and velocity distribution (arrows) on the cross-section in the $y=0$ plane.
Lower panel ($z<0$): Plasma $\beta$ $\betap$ (color-scale, contours) and magnetic field (arrows) on the cross-section in the $y=0$ plane.
}
\label{fig:9}
\end{center}
\end{figure}

To investigate the properties of the outflow and jets,  we show in Figure~\ref{fig:9} the $z$-component of the velocity $v_z$ (upper side) and the plasma beta $\betap$ (lower side) at the same epoch and with the same scale as Figure~\ref{fig:7}{\it d}.
In this figure, we can see two sets of peaks in the $z$-component of the velocity  (outer weak and inner strong peaks; Fig.~\ref{fig:9} upper side).
The outer weak peaks with $v_z\simeq 5\km$ are located at ($x$, $z$) = ($\pm0.3$, 0.3) and originate from the first core.
The inner strong peaks with $v_z \simeq 15\km$ are located at ($x$, $z$) = ($\pm0.05$, 0.3) and originate from the rotation supported quasi-static core inside the first core.
In the upper panel, we label the flow driven from the first core as `outflow', and the flow driven from the quasi-static core as `jet'.
Figure~\ref{fig:9} lower panel shows that the plasma beta is $\betap<1$ around the velocity peak, which indicates that flows with $v_z > 0$ or $v_z < 0$ for $z>0$ or $z<0$, respectively, are driven from the region with strong magnetic field.  
These regions are composed of low-density gas ($\nc \lesssim 10^{13}\cm$), as shown in Figure~\ref{fig:7}{\it d}, and thus Ohmic dissipation is ineffective (Fig.~\ref{fig:1}) and the magnetic field can be amplified as the cloud collapses.
On the other hand, the central object and disk located at $z<\vert 0.1 \vert$ have a large plasma beta ($\betap = 10^5-10^7$) because Ohmic dissipation is effective in this region.
The jet is driven from the second core (protostar) after the protostar formation epoch ($\nc = 10^{21}\cm$) in Model 12, while the jet is driven from the quasi-static core before the protostar formation epoch in Model 16.
This is because the rotation supported core is formed before the central density reaches $\nc = 10^{21}\cm$ in Model 16, due to lower effectiveness of magnetic braking.
Figure~\ref{fig:7}{\it e} shows the cloud structure when the density reaches $\nc = 8.1\times 10^{18}\cm$.
In this figure, a strong jet is driven from the center.

When the gas density reaches $\nc \simeq 1.3 \times 10^{21}\cm$ in Model 16, a shock occurs near the center of the cloud and the protostar (second core) is formed.
The protostar in Model 16 is formed at a lower density than that in Model 12.
The protostar has a mass of $9.1\times 10^{-4}\msun$ and a radius of $3\rsun$ at its formation epoch.
The radius of the protostar in Model 16 is three times larger than that in Model 12. 
The central region inside the first core rotates rapidly in Model 16, thus, a considerably flattened disk (Fig.~\ref{fig:7}{\it f} lower panel) is formed near the center due to the strong centrifugal force.
Figures~\ref{fig:7}{\it f} and \ref{fig:4}{\it f} show that the protostar in Model 16 rotates rapidly and has an oblate structure.
Figure~\ref{fig:8}{\it f} shows that the magnetic field lines are weakly twisted around the protostar, where Ohmic dissipation is effective.
This is because the rapid rotation makes the first core (or high-density region) increasingly oblate as the cloud evolves, and the regions above and below the first core and protostar have low-density gases ($\nc \lesssim 10^{15}\cm$), as shown in Figure~\ref{fig:7}{\it f} lower panel. 
Thus, the magnetic field is decoupled from the neutral gas and again de-collimated in this region.

As a result, Ohmic dissipation affects not only the distribution and strength of the magnetic field but also the cloud collapse, rotation, outflow, and jet.

\subsection{Dependence on $\ceta$}
\begin{figure}
\begin{center}
\includegraphics[width=160mm]{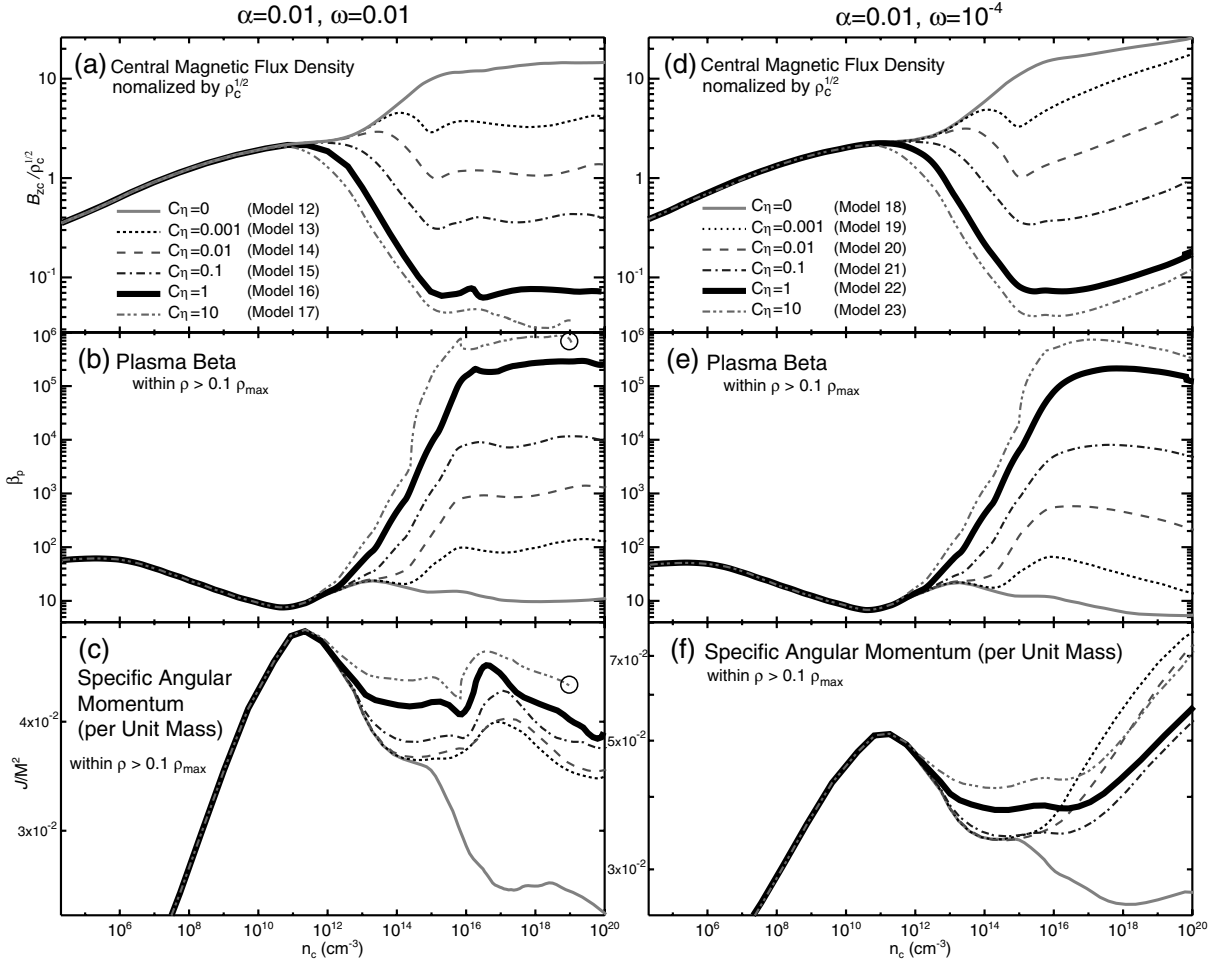}
\caption{
Evolution of models with different diffusivity $\eta$.
Upper panels (a) and (d): Magnetic field $B_{\rm zc}$ normalized by the square root of the density $\rhoc^{1/2}$ in the unit of initial sound speed $c_{s,0}$ against number density at the center of the cloud for (a) Models 12--17 and (d) Models 18--23.
Middle panels (b) and (e): Plasma beta $8\pi c_s^2 \rho/B^2$ within the region $\rho > 0.1 \rho_{\rm c}$ for the same models as panels (a) and (d).
Lower panels (c)  and (f):  $J/M^2$ in the unit of $4\pi G/c_{\rm s,0}$ within $\rho > 0.1 \rho_{\rm c}$ for the same models as panels (a) and (d).
}
\label{fig:6}
\end{center}
\end{figure}

Ohmic dissipation greatly influences the cloud evolution, as described in the previous section.
We estimate the resistivity as a function of the density and temperature according to \citet{nakano02}, shown by the thick solid line in Figure~\ref{fig:1}.
However, we note that it is difficult to determine the ionization structure accurately owing to the uncertainties, particularly in the properties of dust grains that provide the most important sites for recombination in collapsing clouds.
Thus, we parameterize the model of the magnetic dissipation $\ceta$ and investigate the cloud evolution, as shown in Figure~\ref{fig:6}.
Figure~\ref{fig:6} left panels show the evolution of models with the same magnetic field and angular velocity ($\alpha=0.01$, $\omega=0.01$) but different $\ceta$.
In Figure~\ref{fig:6}{\it a}, the larger $\ceta$ efficiently removes the magnetic field from the cloud.
The plasma beta is maintained at $\betap\simeq 10$--20 for $\nc \gtrsim 10^{12}\cm$ in Model 12 (the ideal MHD model; $\ceta=0$), while it reaches $\betap\simeq 10^5$ for $10^{12}\lesssim \nc \lesssim 10^{15}\cm$ in Model 16 ($\ceta=1$) in Figure~\ref{fig:6}{\it b}.
Thus, the central region in Model 12 has magnetic energy $\sim10^4$ times larger than that in Model 16.
This difference in the magnetic energy between Models 12 and 16 is caused by Ohmic dissipation for $10^{12}\cm \lesssim \nc \lesssim 10^{16}\cm$.
The magnetic energy in the central region depends on $\ceta$, as shown in Figure~\ref{fig:6}{\it b}.
In contrast, the angular momentum is large and the cloud rotates rapidly in models with larger $\ceta$, as shown in Figure~\ref{fig:6}{\it c}.
This is because magnetic braking is less effective in models with larger $\ceta$ for less magnetic energy.
As a result, protostars with larger $\ceta$ have larger angular momentum at the formation epoch.
In Model 17 ($\ceta=10$), the cloud rotates rapidly and fragmentation occurs at $\nc \simeq 10^{19}\cm$, because the magnetic energy is small and magnetic braking is barely effective.

Figure~\ref{fig:6} right panels (Fig.~\ref{fig:6}{\it d}--{\it f}) show the evolution of models having the same magnetic field ($\alpha=0.01$) and rotation rate ($\omega = 10^{-4}$) with different $\ceta$.
In the initial state, the rotation rates of the models shown in Figure~\ref{fig:6} right panels are 100 times smaller than those in the left panels ($\omega = 10^{-2}$), while the magnetic field strengths are the same.
The evolution of the magnetic field and plasma beta is almost the same for $\nc \lesssim 10^{15}\cm$ in both Figure~\ref{fig:6}{\it a} and {\it d}.
After the magnetic dissipation, the evolutions of the models shown in the left and right panels are different.
The value $B_{\rm c}/\rhoc^{1/2}$ increases with gas density for $\nc \gtrsim 10^{15}\cm$ in Models 18--23 (Fig.~\ref{fig:6}{\it d}), while it is almost constant in Models 12-17 (Fig.~\ref{fig:6}{\it a}).
The magnetic field is more greatly amplified for a spherical collapse  ($B\propto\rho^{2/3}$) than for a disk-like collapse ($B\propto\rho^{1/2}$), as explained in \S4.1.
In ideal MHD models, even if the cloud has no angular momentum, the geometry of the collapse changes from spherical to disk-like collapse, because the magnetic field is amplified and a disk-like structure is formed at the center of the cloud.
After the disk-like structure is formed (i.e., during the disk-like collapse), the growth rate of the magnetic field becomes small \citep{machida05a}.
However, if the magnetic field is removed from the central region sufficiently, the geometry of the collapse again changes from disk-like to spherical collapse and the magnetic field is amplified as $B\propto\rho^{2/3}$ (spherical collapse).
In Models 18--21, after the dissipation of the magnetic field ($\nc \gtrsim 10^{16}\cm$), the clouds collapse spherically because of the extremely slow rotation, while in Models 12--17 the clouds collapse with a disk-like shape due to the rapid rotation.
At the protostar formation epoch ($\nc \simeq 10^{21}\cm$), the magnetic field is amplified to $B_{\rm c} = 0.11\kg$ in Model 16 (rapidly rotating cloud), while it is amplified to  $B_{\rm c}= 0.68\kg$ in Model 22 (slowly rotating cloud).
Thus, the protostar has a stronger magnetic field for a more slowly rotating cloud, even in non-ideal MHD models.
After the magnetic field is significantly removed from the center, the angular velocity is also amplified because the cloud collapses spherically.
Thus, the specific angular momentum normalized by the mass $J/M^2$ also increases in Models 18--23 (Fig.~\ref{fig:6}{\it f}), while $J/M^2$ is constant in the more rapidly rotating clouds of Models 12--17 for disk-like collapse (Fig.~\ref{fig:6}{\it c}).

\subsection{Cloud Evolution with Different Angular Momentum}
\begin{figure}
\begin{center}
\includegraphics[width=130mm]{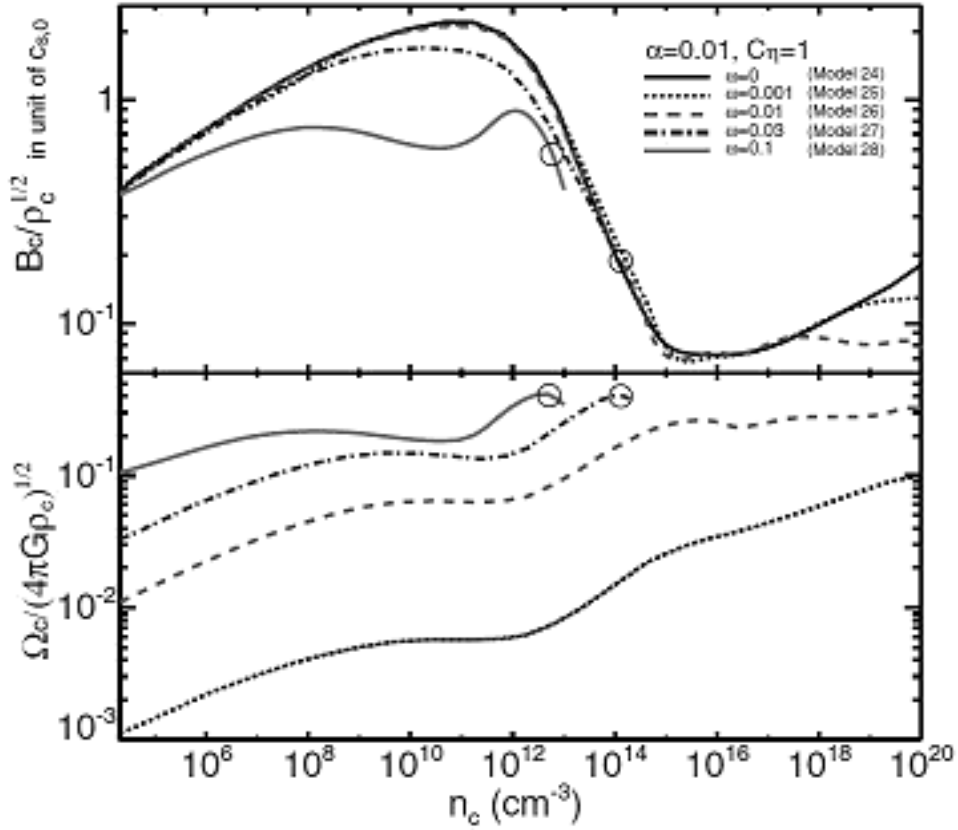}
\caption{
Evolutions of the magnetic field and angular velocity for various initial rotation rates.
Upper panel: Magnetic field $B_{\rm c}$ normalized by the square root of the density $\rhoc^{1/2}$ in the unit of initial sound speed $c_{s,0}$ against the number density at the centers of the clouds for Models 24--28.
Lower panel: Angular velocity $\Omega_{\rm c}$ normalized by the free-fall rate [$(4\pi G \rhoc)^{1/2}$] at the centers of the clouds for Models 24--28.
}
\label{fig:10}
\end{center}
\end{figure}

Figure~\ref{fig:10} shows the evolution of the magnetic field $B_{\rm c}/\rhoc^{1/2}$ (upper panel) and angular velocity $\Omega/(4\pi G \rhoc)^{1/2}$ (lower panel) for Models 24--28.
These models have the same magnetic field ($\alpha=0.01$) and $\ceta$ ($\ceta=1$), but different angular velocities $\omega$.
In this figure, the circles indicate the fragmentation epoch.
Fragmentation occurs in the rapidly rotating clouds of Models 27 ($\omega=0.03$) and 28 ($\omega=0.1$).
\citet{machida05b} shows that fragmentation occurs when the central angular velocity $\Omega_{\rm c}$ normalized by the free-fall timescale $1/(4\pi\,G\,\rhoc)^{1/2}$  reaches $\Omega_{\rm c}/(4\pi{\rm G}\rho)^{1/2} \simeq 0.2$ for the isothermal phase $\nc \lesssim 10^{11}\cm$.
As shown in Figure~\ref{fig:10} lower panel, this condition is realized in Models 27 and 28.
After fragmentation occurs, we stopped the calculation because our numerical code (nested grid) is not suitable for calculating the evolution of each fragment located outside the center.
The protostar forms without fragmentation in Models 24, 25, and 26, because the fragmentation condition is not realized in these clouds, as shown in Figure~\ref{fig:10} lower panel.

Figure~\ref{fig:10} upper panel shows the evolution of the magnetic field $B_{\rm c}/\rhoc^{1/2}$.
In this panel, the clouds without fragmentation have almost the same magnetic field strength for $\nc \lesssim 10^{16}\cm$.
For $\nc \gtrsim 10^{16}\cm$, the magnetic field continues to increase in the non-rotating cloud (Model 24), while they saturate at certain values in the rotating clouds (Models 25 and 26).
A close analysis of the evolutions for $\nc \gtrsim 10^{16}\cm$ shows that the growth rate of the magnetic field becomes small when the normalized angular velocity approaches $\Omega_{\rm c}/(4\pi G \rhoc)^{1/2} \simeq 0.2$.
This is due to the fact that the collapse of the geometry changes from spherical to disk-like when the rotational energy becomes comparable to the gravitational energy [for details, see \citealt{machida05a}].
The protostars have magnetic fields of $B_{\rm c}= 0.51\kg$ (Model 24; $\omega =0$), $0.48\kg$ (Model 25; $\omega=0.001$), and $0.11\kg$ (Model 26; $\omega=0.01$).
In this way, the magnetic field strengths of formed protostars are related to the initial cloud rotation.
In summary, a protostar has a stronger magnetic field in a more slowly rotating cloud.

Observations indicate that molecular clouds have large amounts of magnetic energy \citep{crutcher99} and small amounts of rotational energy \citep{goodman93,caselli02}.
These clouds are classified as magnetic-force dominant clouds ($\Omega_0/B_0 < \Omega_{\rm cri}/B_{\rm cri} \equiv 0.39\, G^{1/2} \, c_{s,0}^{-1}$.) in the criterion of \citet{machida05a,machida06a}.
As shown in \S4.1, the magnetic field converges to a certain value for the isothermal phase in magnetic-force dominant clouds.
After the magnetic dissipation, the growth rate of the magnetic field depends on the rotation rate at the center of the cloud.   
Thus, the magnetic field of a protostar is determined almost completely by the rotation energy of the initial cloud when the degree of resistivity $\ceta$ is fixed.

\section{Discussion}
\subsection{Magnetic Flux Problem}
The magnetic flux problem is important for the star formation process.
Magnetic energy in an interstellar cloud is usually regarded as being comparable to the gravitational energy of the cloud, while the magnetic energy in a star is much smaller than the gravitational energy.
The molecular cloud can collapse when the following condition is realized:
\begin{equation}
\Phi < \Phi_{\rm cri},
\end{equation}
where $\Phi$ is the magnetic flux and is defined as 
\begin{equation}
\Phi = \int \vect{B}\cdot d\vect{S},
\end{equation}
and $\Phi_{\rm cri}$ is the critical value of the collapse and is defined as
\begin{equation}
\Phi_{\rm cri} = f G^{1/2} M,
\end{equation}
where $f$ is a dimensionless constant ($f\approx8$; \citealt{mouschovias76,tomisaka88}) and $M$ is the mass of the cloud or cloud core\footnote{
In general, $\Phi$ and $\Phi_{\rm cri}$ are used to measure the degree of the magnetic field strength of an isolated cloud core.
In this paper, however, we use these values to measure the degree of the magnetic field strength inside a collapsing cloud core.
We define $\Phi$ and $\Phi_{\rm cri}$ simply as functions of the magnetic field and mass in the local region inside the cloud core.
}.
The observed magnetic fluxes of molecular cloud cores \citep[e.g.,][]{crutcher99} are close to their critical value ($\Phi/\Phi_{\rm cri}\sim1$), while those of protostars are much smaller than their critical value ($\Phi/\Phi_{\rm cri}\approx 10^{-5}$--$10^{-3}$; \citealt{johns01}).
Thus, the magnetic flux should be largely removed from the cloud core during the star formation process.
This is called ``the magnetic flux problem'' in star formation.

The flux to mass ratio $\Phi/M$ around the center of a collapsing cloud decreases by two mechanisms. 
One is vertical collapse \citep{nakano83, machida05a} and the other is Ohmic dissipation \citep{nakano02}.
When a gas infalls vertically along the magnetic field line (vertical collapse), the mass around the central region increases, keeping the magnetic flux constant.
Thus, the ratio of the magnetic flux to the critical mass, $\Phi/\Phi_{\rm cri}$ ($\propto \Phi/M$), decreases,  because the denominator of $\Phi/\Phi_{\rm cri}$ increases, while the numerator is nearly constant.
However, \citet{nakano83} showed that the vertical collapse mechanism is not sufficient for solving the magnetic flux problem, because a large amount of gas collects from a wide range ($>200$\,kpc) to decrease this ratio from $\Phi/\Phi_{\rm cri}\simeq 1$ to $ \sim 10^{-5}$--$10^{-3}$.
On the other hand,  \citet{nakano02} showed analytically that the Ohmic dissipation process in a collapsing cloud is an effective mechanism for solving the magnetic flux problem.
We numerically confirmed that Ohmic dissipation is more important than vertical collapse for the removal (or decrease) of magnetic flux in a collapsing cloud.

\begin{figure}
\begin{center}
\includegraphics[width=150mm]{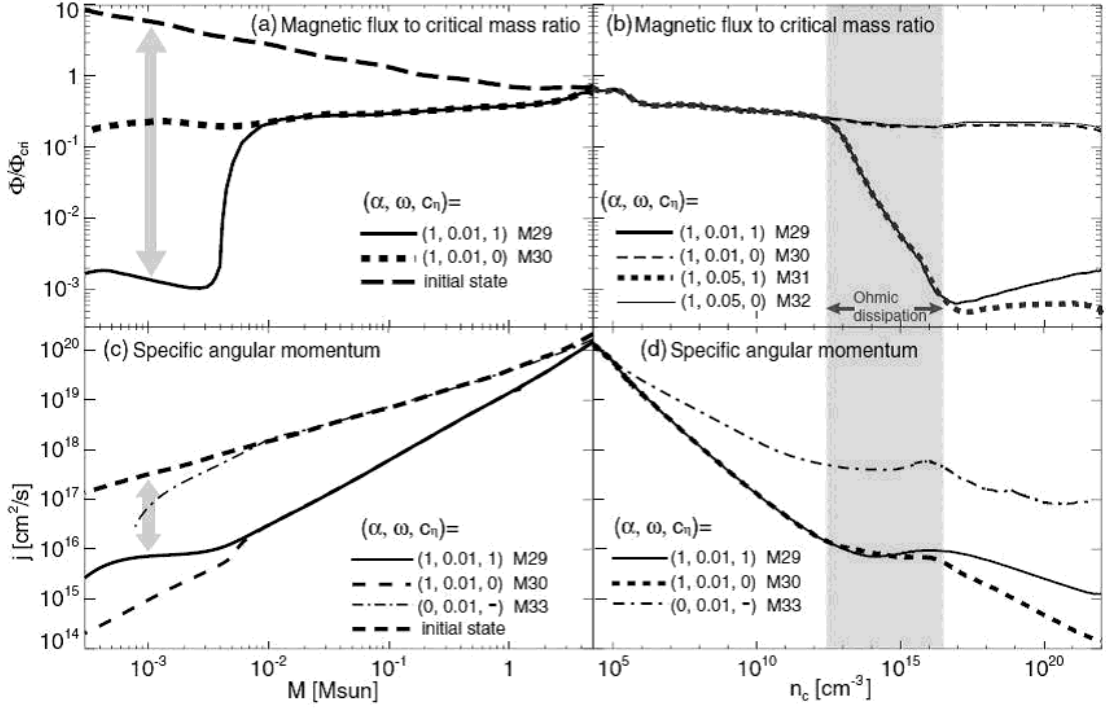}
\caption{
(a) Spatial distribution of magnetic flux. 
The magnetic flux normalized by the critical mass $\Phi/\Phi_{\rm cri}$ as a function of the cumulative mass from the center of the cloud at the initial state (broken line) and at the end of the calculation for Models 29 (solid line) and 30 (dotted line) are plotted.
(b) Evolution of $\Phi/\Phi_{\rm cri}$ within $\rho > 0.1 \rhoc$ against the central number density for Models 29--32. 
(c) Spatial distribution of specific angular momentum.
Specific angular momentum $j$ as a function of the cumulative mass from the center of the cloud at the initial state (thick broken line) and the end of the calculation for Models 29 (solid line), 30 (thin broken line), and 33 (dash-dotted line) are plotted.
(d) Evolution of $j$  within $\rho > 0.1 \rhoc$ against the number density at the center of the cloud for Models 29, 30, and 33.
}
\label{fig:11}
\end{center}
\end{figure}

Figure~\ref{fig:11}{\it a} shows the distribution of the magnetic flux normalized by the critical mass $\Phi/\Phi_{\rm cri}$ as a function of the cumulative mass from the center of the cloud at the initial state (broken line), and at the end of the calculation for Models 29 (solid line) and 30 (dotted line).
In this graph, the mass $M$ ($x$-axis) is integrated for every iso-density contour $\rho_{\rm a}$ from the center that has maximum density, as
\begin{equation}
M (\rho > \rho_{\rm a}) = \int_{\rho >  \rho_{\rm a}} \rho \, dV,
\end{equation}
and the magnetic flux inside the corresponding region is integrated as
\begin{equation}
\Phi (\rho > \rho_{\rm a})  = \int_{\rho > \rho_{\rm a}, z=0} \vect{B}\cdot d\vect{S},
\end{equation}
where $d\vect{S}$ is defined at the $z=0$ plane and parallel to the $z$-axis.
The graph indicates that Models 29 and 30 have magnetically supercritical clouds ($\Phi/\Phi_{\rm cri}<1$) as a whole ($M>0.3\msun$), while the central regions ($M<0.3\msun$) of these clouds are magnetically subcritical ($\Phi/\Phi_{\rm cri}<1$) at the initial state. 
The ratio $\Phi/\Phi_{\rm cri}$ increases with decreasing mass.
Since we adopt the Bonnor--Ebert density profile \citep{bonnor56,ebert55}, the initial density distribution near the center is almost constant and thus the mass of the cloud (i.e., $\Phi_{\rm cri}$) is proportional to $R^3$, $M\propto \Phi_{\rm cri}\propto R^3 $.
The magnetic flux is proportional to $\Phi \propto R^2$, because a uniform magnetic field is assumed at the initial state.
Thus, the central region (comprising a small fraction of the total mass) has large  $\Phi/\Phi_{\rm cri}$, since the ratio of the magnetic flux to the critical mass is proportional to $\Phi/\Phi_{\rm cri} \propto R^{-1} \propto M^{-1/3}$ near the center.
However, these clouds can collapse promptly, because a large amount of the mass ($\sim 99\%$ of the total mass) is magnetically supercritical.

In our calculations, protostars have a mass of $\sim$$10^{-3}\msun$ at their formation epoch.
Figure~\ref{fig:11}{\it a} shows that the ratio of the magnetic flux to the critical mass within a mass of $M=10^{-3}\msun$ (i.e., at $M=10^{-3}\msun$ on the $x$-axis) decreases from $\Phi/\Phi_{\rm cri} =6$ to $\Phi/\Phi_{\rm cri}\simeq 0.2$  in Model 30 (ideal MHD model).
In Model 30, the ratio $\Phi/\Phi_{\rm cri}$ decreases due to vertical collapse, because Ohmic dissipation is ignored since $\ceta=0$.
This indicates that the magnetic flux is not removed sufficiently by the vertical collapse, as shown in \citet{nakano83}.
On the other hand, in Model 29 (non-ideal MHD model), the ratio $\Phi/\Phi_{\rm cri}$  within $M=10^{-3}\msun$ decreases from $\Phi/\Phi_{\rm cri} =6$ to $\Phi/\Phi_{\rm cri}\simeq 10^{-3}$ (Fig.~\ref{fig:11}{\it a}).
Thus, three or four orders of magnitude of the initial magnetic flux are removed from the core by  Ohmic dissipation in Model 29. 
The solid line in Figure~\ref{fig:11}{\it a} shows that the ratio $\Phi/\Phi_{\rm cri}$ decreases from $\Phi/\Phi_{\rm cri}\simeq 0.1$ to $\Phi/\Phi_{\rm cri}\simeq 10^{-3}$ steeply between  $10^{-2} \msun \lesssim M \lesssim 5\times 10^{-3}\msun$.
This drop indicates that the gas within $M < 10^{-2}\msun$ from the center of the cloud experiences Ohmic dissipation.
The remainder of the cloud is expected to experience Ohmic dissipation as the collapse continues.

Figure~\ref{fig:11}{\it b} shows the evolution of $\Phi/\Phi_{\rm cri}$ for Models 29--32 against the central density. 
In this graph, the magnetic flux $\Phi$ and the mass $M$ within the gas having density $ \rho > 0.1 \rhoc$ are integrated at each timestep as
\begin{equation}
M (\rho > 0.1 \rhoc) = \int_{\rho > 0.1 \rhoc} \rho \, dV,
\label{eq:mass}
\end{equation}
\begin{equation}
\Phi (\rho > 0.1 \rhoc) = \int_{\rho > 0.1 \rhoc, z=0} \vect{B} \cdot d\vect{S}.
\end{equation} 
$\Phi/\Phi_{\rm cri}$ in Models 30 and 32 (ideal MHD models) decreases slightly from the initial state ($\Phi/\Phi_{\rm cri}\simeq 0.8$) to the protostar formation epoch ($\Phi/\Phi_{\rm cri}\simeq 0.2$).
In contrast, for Models 29 and 31 (non-ideal MHD models), $\Phi/\Phi_{\rm cri}$  decreases significantly for $2\times 10^{12}\cm \lesssim \nc \lesssim 6 \times 10^{15}\cm$ by Ohmic dissipation, and reaches $\Phi/\Phi_{\rm cri} \simeq 10^{-3}$ at the protostar formation epoch.

In the non-ideal MHD models, the formed protostars have smaller $\Phi/\Phi_{\rm cri}$ ratios for the more rapidly rotating clouds.
After the magnetic dissipation ($\nc \gtrsim 10^{16}\cm$), the evolution of $\Phi/\Phi_{\rm cri}$ is different in the non-ideal Models 29 and 31 (Fig.~\ref{fig:11}{b}). 
For this phase ($\nc \gtrsim 10^{16}\cm$), the ratio is almost constant ($\Phi/\Phi_{\rm cri}\simeq 5\times 10^{-4}$) in Model 31 (with a rapidly rotating cloud), while it continues to increase in Model 29 (with a slowly rotating cloud).
At the protostar formation epoch, the protostars have $\Phi/\Phi_{\rm cri}=2\times 10^{-3}$ for Model 29 and $\Phi/\Phi_{\rm cri}=4\times 10^{-4}$ for Model 31.
This difference is caused by the rotation rate of the initial cloud.
In Model 31, the cloud collapses vertically (i.e., vertical collapse) and the growth rate of the magnetic field is low because of the rapid rotation for $\nc \gtrsim 10^{16}\cm$.
On the other hand, in Model 29, the cloud collapses spherically (i.e., spherical collapse) and the growth rate of the magnetic field is high for $\nc \gtrsim 10^{16}\cm$, because both the Lorentz and centrifugal forces are weak. 
After the magnetic dissipation ($\nc \gtrsim 10^{16}$), since the Lorentz force is weak, the geometry of the collapse is determined only by the cloud rotation.
An initially slowly rotating cloud collapses spherically to form a protostar with a large magnetic flux,  while a rapidly rotating cloud collapses along the rotation axis to form a protostar with a small magnetic flux.

In summary, protostars have $\Phi/\Phi_{\rm cri}\simeq 10^{-4}$ to $10^{-3}$ at their formation epoch, which is comparable to observations.
However, since we stopped our calculation at $M_{\rm protostar} \sim 10^{-3}\msun$, further calculations are needed to predict the magnetic flux for protostars with $M_{\rm protostar} \sim 1 \msun$.

\subsection{Angular Momentum Problem}
The angular momentum problem is also important for the star formation process.
Molecular cloud cores have specific angular momenta of the order of $j_{\rm cloud} \approx 10^{21} \jcm$ \citep{goodman93,caselli02}, while protostars have specific angular momenta of the order of $j_{\rm protostar} \approx 10^{16}\jcm$.
Thus, the specific angular momentum of a protostar is $\sim 10^{-5}$ times smaller than that of a molecular cloud core. 
For this reason, the angular momentum of the initial cloud must be removed from the core as well as the magnetic flux.
\citet{tomisaka00,tomisaka02} calculated the cloud evolution from a molecular cloud to protostar formation paying attention to the evolution of the angular momentum in two-dimensional nested grid simulations.
The study showed that the angular momentum of the initial cloud is sufficiently removed by magnetic braking and outflow in the collapsing cloud, and the formed protostar has a specific angular momentum comparable to observed values.
However, the ideal MHD approximation was used and thus the magnetic effect (i.e., magnetic braking and outflow) may have been overestimated.
In this section, we discuss the evolution of the angular momentum in the non-ideal MHD regime.

Figure~\ref{fig:11}{\it c} shows the distributions of the specific angular momentum against the cumulative mass at the initial state (thick broken line), and the end of the calculation for Models 29 (solid line), 30 (thin broken line), and 33 (dash-dotted lines). 
In this figure,  the specific angular momentum is integrated up to every iso-density surface $\rho_{\rm a}$ from the center as
\begin{equation}
j(\rho > \rho_{\rm a}) = \dfrac{1}{M(\rho > \rho_{\rm a})}  \int_{\rho > \rho_{\rm a}} \rho \, \varpi \cdot v_{\phi}  \, dV,
\end{equation}
where $M (\rho >  \rho_{\rm a})$ is the mass within $\rho > \rho_{\rm a}$ [see Eq.~(\ref{eq:mass})] and $\varpi$ is the radius in cylindrical coordinates.
As shown by the thick dashed line, the specific angular momentum at the initial state is proportional to $j_{\rm cloud} \propto R^2$ near the center of the cloud, starting with rigid rotation and the Bonnor--Ebert density profile with an almost constant density at the center.

In Figure~\ref{fig:11}{\it c}, the specific angular momentum at the end of the calculation ($j_{\rm protostar}$) for Model 33 is shown as a dash-dotted line. 
In Model 33, the specific angular momentum is not removed by magnetic effects (magnetic braking, outflow and jet), because the cloud is unmagnetized.
Thus, the distribution of the specific angular momentum at the end of the calculation is almost the same as that of the initial state for $M>10^{-2}\msun$.
This indicates that the angular momentum is almost conserved outside the protostar ($M>10^{-2}\msun$). 
For $M<10^{-2}\msun$, the specific angular momentum at the end of the calculation is slightly smaller than that at the initial state, caused by angular momentum transfer by the non-axisymmetric pattern.
In this model, after a thin disk is formed, the disk is transformed into a bar, because bar mode instability \citep[e.g.,][]{durisen86} is caused by rapid rotation.
The specific angular momentum begins to decrease around the center of the cloud as the non-axisymmetric pattern grows.
Fragmentation in this unmagnetized model is observed for further calculations, as also shown by \citet{banerjee06}.
As discussed in \citet{machida05b}, fragmentation plays an important role in the angular momentum evolution, because the angular momentum is redistributed into the orbital and spin angular momentum of each fragment.
We will discuss the fragmentation and redistribution of the angular momentum in subsequent papers.

The specific angular momenta of magnetized clouds at the end of the calculation are plotted by the solid (non-ideal MHD Model 29) and thin broken lines (ideal MHD Model 30) in Figure~\ref{fig:11}{\it c}.
The specific angular momenta of both Models 29 and 30 at the end of the calculation are smaller than those at the initial state.
In these models, the specific angular momentum decreases by vertical collapse, magnetic braking, outflow, and jets.
When the initial cloud is magnetized strongly (or rotates rapidly), the cloud collapses vertically along the magnetic field line (i.e., vertical collapse).
Gas with small specific angular momentum preferentially falls into the central region, and thus the central region gains specific angular momentum in the vertical collapse regime.
Note that vertical collapse does not transfer angular momentum, because this mechanism changes only the location of each fluid element and thus the gas keeps its specific angular momentum. 
For higher densities ($\nc \gtrsim 10^{11}\cm$), the magnetic field lines are strongly twisted, and outflow and jets appear.
The angular momentum is effectively transferred by the outflow and jets, as already shown by \citet{tomisaka00,tomisaka02}.
Magnetic braking is also important for the angular momentum transfer.
In principle, however, a reduction of the angular momentum due to magnetic braking cannot be distinguished from that due to outflow and jets, because outflow and jets are consequences of the magnetic braking process.

Figure~\ref{fig:11}{\it d} shows the evolution of the specific angular momentum $j$ in the inner region for Models 29, 30, and 33 as a function of the central density. 
The angular momentum is integrated within $\rho > 0.1\rho_{\rm c}$ at each timestep and averaged as
\begin{equation}
j (\rho > 0.1 \rhoc) = \dfrac{1}{M(\rho > 0.1 \rhoc)}  \int_{\rho > 0.1 \rho_{\rm c}} \rho \, \varpi \cdot v_{\phi}  \, dV,
\end{equation}
where $M(\rho > 0.1 \rhoc)$ is the mass within $\rho > 0.1 \rho_{\rm c}$.
In this graph, the specific angular momenta in the magnetized clouds (Models 29 and 30) are smaller than that in the unmagnetized cloud (Model 33) for $n_c \gtrsim 10^{6}\cm$.
The difference of the specific angular momentum for the isothermal phase ($\nc \lesssim 10^{11}\cm$) is due to magnetic braking and vertical collapse, because no outflow appears in this phase.
The specific angular momentum in Model 30 (ideal MHD model) is smaller than that in Model 29 (non-ideal MHD model) for $\nc \gtrsim 10^{15}\cm$, because magnetic braking is less effective due to magnetic dissipation for $2\times 10^{12}\cm \lesssim \nc \lesssim 6 \times 10^{15}\cm$.

At the protostar formation epoch ($\nc = 10^{21}\cm$), protostars have specific angular momenta of $j_{\rm protostar} = 8\times 10^{16}\,\jcm$ for Model 31, $1.5\times 10^{15}\,\jcm$ for Model 29, and $2\times 10^{14}\,\jcm$ for Model 30.
Observations indicate that classical T Tauri stars have specific angular momenta of the order of $j_{\rm obs}\sim 10^{16}\jcm$ \citep{bouvier93,herbst06}.
At the end of our calculation, the formed protostars have masses of only $\sim 10^{-3}\msun$.
Since the spatial distribution of the specific angular momentum follows $j\propto \varpi^2$ for rigid rotation, the region inside $M < 10^{-3}\msun$ of T Tauri stars has an average specific angular momentum of $\sim$$10^{14}\jcm$.
Thus, the specific angular momenta in non-ideal MHD models such as Model 29 are about one order of magnitude larger than those of observed protostars.
In this paper, we discuss the angular momentum evolution only in the early star formation phase ($M_{\rm protostar}<10^{-2}-10^{-3}\msun$).
In the later accretion phase, however, the angular momentum is transferred by star--disk interaction, stellar wind, and jets.
To fully solve the angular momentum problem, the evolution in the later phase must also be calculated.

This paper focuses on the formation of single stars, and hence we have investigated the evolution of slowly rotating clouds.
As shown in \citet{cha03}, \citet{matsumoto03}, and \citet{machida05b}, fragmentation occurs in a collapsing cloud core and binary or multiple stars are formed in initially rapidly rotating clouds.
When binary or multiple stars are formed, the angular momentum is redistributed in the orbital and spin angular momenta of each star \citep{machida05b}.
Thus, the binary or multiple star formation process is also important for the angular momentum problem.
We will investigate the evolution in initially rapidly rotating clouds in subsequent papers.

\subsection{Comparison with Observations}
In our calculation, the formed protostars have magnetic fields of $B_{\rm protostar} = 0.16\kg$ (Model 16) to $1.53\kg$ (Model 32) in models with $\eta=1$ (non-ideal MHD Models 16, 22, 24, 25, 26, 27, and 28).
On the other hand, a surface magnetic field of the order of 1--3$\kg$ has been derived from Zeeman broadening measurements of CTTS photospheric lines \citep{johns99a,johns99b, johns01, guenther99,bouvier05}.
Thus, the magnetic fields derived from our calculation are consistent with observations. 
On the other hand, the rotation periods of protostars in our calculation are slightly shorter than the observations.
The formed protostars have rotation periods of $P=0.15$ days (Model 26) to 1.53 days (Model 32) in models with $\ceta=1$, while observed protostars have rotation periods of $P=1$--10 days \citep{herbst06}.

Since our calculations show only the early star formation phase, in which the protostars have masses of $\simeq 10^{-3}$--$10^{-2} \msun$.
The magnetic field strength and angular momentum of a protostar may change in subsequent gas accretion phase.
The magnetic field may be amplified by magneto-rotational instability \citep{balbus91} and by shearing motion between the star and the ambient medium.
The angular momentum of a protostar may increase due to accretion of gas with large angular momentum, and decrease by jets or star--disk interaction \citep[e.g.,][]{matt05a,matt05b}.
To fully describe the magnetic field and angular momentum of protostars, we must calculate the cloud evolution up to the later accretion phase.
However, our results indicate that at its formation epoch, a protostar already has a magnetic field strength and rotation period almost of the same order of magnitude as observations.
Thus, the magnetic field and angular momentum of observed protostars may be largely determined by the early star formation phase before the main accretion phase.

\acknowledgments
We have greatly benefited from discussions with ~T. Nakano, and ~K. Saigo.
We also thank T. Hanawa for his contribution to the nested grid code.
Numerical calculations were carried out with a Fujitsu VPP5000 at the Astronomical Data Analysis Center, National Astronomical Observatory of Japan.
This work is supported by a Grant-in-Aid for the 21st Century COE ``Center for Diversity and Universality in Physics'' from the Ministry of Education, Culture, Sports, Science and Technology (MEXT) of Japan, and partially supported by 
the Grants-in-Aid from MEXT (15740118, 16077202, 18740113,18740104).

\begin{table}   
\setlength{\tabcolsep}{2pt}
\caption{Model parameters and calculation results}
\label{table:init}
\begin{center}
\begin{tabular}{ccccccccccccccccccc|cc}
\hline
 Model & $\alpha$  & $\omega$ & $\ceta$ & $f$ &  $B_0$ {\scriptsize [$\mu$G]} & $\Omega_0$ {\scriptsize [s$^{-1}$]}  &  $M${\scriptsize [$\msun$]} &$\alpha_0$ &$\beta_0$ & $\gamma_0$ & $B_{\rm f}$\, {\scriptsize (k\,G)}  & P\, {\scriptsize (day)} \\
\hline
1 & 5        & 0 & 0 & 1.4& 34 & 0 & 6.3 & 0.6 & 0 & 5.8  & 89.7 & $\infty$ \\
2 & 0.5      & 0 & 0 & 1.4& 11 & 0 & 6.3 & 0.6 & 0 & 0.58 & 123 & $\infty$ \\
3 & 0.1     & 0 & 0 & 1.4 & 5.0& 0 & 6.3 & 0.6 & 0 & 0.12 & 106 & $\infty$ \\
4 & 0.05     & 0 & 0 & 1.4& 3.5& 0 & 6.3 & 0.6 & 0 &$5.8 \times 10^{-2}$  & 102 & $\infty$ \\
5 & 0.01     & 0 & 0 & 1.4& 1.6& 0 & 6.3 & 0.6 & 0 &   $1.2 \times 10^{-2}$ & 92.5 & $\infty$ \\
6 & $10^{-3}$& 0 & 0 & 1.4& 0.5& 0 & 6.3 & 0.6 & 0 & $1.2 \times 10^{-3}$ & 88.9 & $\infty$ \\
\hline
7 & 0.01& $10^{-3}$& 0 & 1.4 &  1.6 &$2.1\times 10^{-16}$ & 6.3& 0.6 & $3.3\times10^{-6}$ & $1.2 \times 10^{-2}$ &95.9& 46.8\\
8 & 0.01& 0.01     & 0 & 1.4 &  1.6 &$2.1\times 10^{-15}$ & 6.3& 0.6 & $3.3\times10^{-4}$ & $1.2 \times 10^{-2}$ &67.3& 5.68\\
9 & 0.01& 0.05     & 0 & 1.4 &  1.6 &$1.1\times 10^{-14}$ & 6.3& 0.6 & $8.2\times10^{-3}$ & $1.2 \times 10^{-2}$ &46.2& 4.30\\
10 & 0.01& 0.1      & 0 & 1.4 &  1.6 &$2.1\times 10^{-14}$ & 6.3& 0.6 & $3.3\times10^{-2}$ & $1.2 \times 10^{-2}$ &43.7& 2.47\\
11 & 0.01& 0.15     & 0 & 1.4 &  1.6 &$3.1\times 10^{-14}$ & 6.3& 0.6 & $7.4\times10^{-2}$ & $1.2 \times 10^{-2}$ &---&  ---\\
\hline
12 & 0.01& 0.01 & 0        & 1.2& 1.4& $2.0\times10^{-15}$ & 5.4 & 0.7 & $3.3\times10^{-4}$ & $1.2 \times 10^{-2}$&42.7& 4.79\\
13 & 0.01& 0.01 & $10^{-3}$& 1.2& 1.4& $2.0\times10^{-15}$ & 5.4 & 0.7 & $3.3\times10^{-4}$ & $1.2 \times 10^{-2}$&2.74& 3.99\\
14 & 0.01& 0.01 & 0.01     & 1.2& 1.4& $2.0\times10^{-15}$ & 5.4 & 0.7 & $3.3\times10^{-4}$ & $1.2 \times 10^{-2}$&0.93& 3.89\\
15 & 0.01& 0.01 & 0.1      & 1.2& 1.4& $2.0\times10^{-15}$ & 5.4 & 0.7 & $3.3\times10^{-4}$ & $1.2 \times 10^{-2}$&0.64& 3.77\\
16 & 0.01& 0.01 & 1        & 1.2& 1.4& $2.0\times10^{-15}$ & 5.4 & 0.7 & $3.3\times10^{-4}$ & $1.2 \times 10^{-2}$&0.11& 3.68\\
17 & 0.01& 0.01 & 10       & 1.2& 1.4& $2.0\times10^{-15}$ & 5.4 & 0.7 & $3.3\times10^{-4}$ & $1.2 \times 10^{-2}$&---& ---\\
\hline
18 & 0.01& $10^{-4}$ & 0        & 1.2& 1.4& $2.0\times10^{-17}$& 5.4& 0.7& $3.3\times10^{-8}$ & $1.3 \times 10^{-2}$&60.9& 271.1\\
19 & 0.01& $10^{-4}$ & $10^{-3}$& 1.2& 1.4& $2.0\times10^{-17}$& 5.4& 0.7& $3.3\times10^{-8}$ & $1.3 \times 10^{-2}$&41.7& 81.2\\
20 & 0.01& $10^{-4}$ & 0.01     & 1.2& 1.4& $2.0\times10^{-17}$& 5.4& 0.7& $3.3\times10^{-8}$ & $1.3 \times 10^{-2}$&13.2&72.6\\
21 & 0.01& $10^{-4}$ & 0.1      & 1.2& 1.4& $2.0\times10^{-17}$& 5.4& 0.7& $3.3\times10^{-8}$ & $1.3 \times 10^{-2}$&2.81&102\\
22 & 0.01& $10^{-4}$ & 1        & 1.2& 1.4& $2.0\times10^{-17}$& 5.4& 0.7& $3.3\times10^{-8}$ & $1.3 \times 10^{-2}$&0.68&109\\
23 & 0.01& $10^{-4}$ & 10       & 1.2& 1.4& $2.0\times10^{-17}$& 5.4& 0.7& $3.3\times10^{-8}$ & $1.3 \times 10^{-2}$&0.32&72.9\\
\hline
24 & 0.01& $ 0 $  & 1 & 1.2&  1.4& 0                  & 5.4& 0.7&                  0 & $1.3 \times 10^{-2}$&0.51&$\infty$ \\
25 & 0.01& $0.001$& 1 & 1.2&  1.4& $2.0\times10^{-16}$& 5.4& 0.7& $3.3\times10^{-6}$ & $1.3 \times 10^{-2}$&0.48& 11.1\\
26 & 0.01& $0.01$ & 1 & 1.2&  1.4& $2.0\times10^{-15}$& 5.4& 0.7& $3.3\times10^{-4}$ & $1.3 \times 10^{-2}$&0.11& 4.21\\
27 & 0.01& $0.03$ & 1 & 1.2&  1.4& $5.9\times10^{-15}$& 5.4& 0.7& $3.0\times10^{-3}$ & $1.3 \times 10^{-2}$&--- &--- \\
28 & 0.01& $0.1$  & 1 & 1.2&  1.4& $2.0\times10^{-14}$& 5.4& 0.7& $3.3\times10^{-2}$ & $1.3 \times 10^{-2}$&--- &--- \\
\hline
29 & 1& 0.01& 1 & 1.4&  16& $2.1\times10^{-15}$& 6.3& 0.6& $3.3\times10^{-4}$ & 1.2&0.89& 5.56\\
30 & 1& 0.01& 0 & 1.4&  16& $2.1\times10^{-15}$& 6.3& 0.6& $3.3\times10^{-4}$ & 1.2&36.1 &59.9 \\
31 & 1& 0.05& 1 & 1.4&  16& $1.3\times10^{-14}$& 6.3& 0.6& $3.3\times10^{-4}$ & 1.2&0.30& 5.38\\
32 & 1& 0.05& 0 & 1.4&  16& $1.3\times10^{-14}$& 6.3& 0.6& $3.3\times10^{-4}$ & 1.2&28.4 &12.0 \\
33 & 0& 0.01& --- & 1.4&0& $2.1\times10^{-15}$& 6.3& 0.6& $3.3\times10^{-4}$ & 1.2&--- &4.86 \\

\hline
\end{tabular}
\end{center}
\end{table}


\clearpage

\clearpage

\clearpage

\clearpage

\clearpage

\clearpage

\clearpage

\clearpage

\end{document}